\documentclass[longauth]{aa} 

\usepackage{color}

\newcommand{\kms}{km~s$^{-1}$}

\newcommand{\vsini}{$v_{\mathrm{eq}}\sin{i_\star}$}

\usepackage{graphicx}
\usepackage{txfonts}
\usepackage{hyperref}
\usepackage{longtable}

\begin{document} 

   \title{The GAPS programme at TNG\fnmsep\thanks{Based on observations made with the Italian Telescopio Nazionale Galileo (TNG) operated by the Fundaci\'{o}n Galileo Galilei (FGG) of the Istituto Nazionale di Astrofisica (INAF) at the Observatorio del Roque de los Muchachos (La Palma, Canary Islands, Spain).}}

   \subtitle{XLIV. Projected rotational velocities of 273 exoplanet-host stars observed with HARPS-N}

   \author{M. Rainer
          \inst{1}
          \and S. Desidera\inst{2}
          \and F. Borsa\inst{1}
          \and D. Barbato\inst{3,4}
          \and K. Biazzo\inst{5}
          \and A. Bonomo\inst{3}
          \and R. Gratton\inst{2}
          \and S. Messina\inst{6}
          \and G. Scandariato\inst{6}
          \and L. Affer\inst{7}
          \and S. Benatti\inst{7}
          \and I. Carleo\inst{8}
          \and L. Cabona\inst{2}
          \and E. Covino\inst{9}
          \and A.F. Lanza\inst{6}
          \and R. Ligi\inst{10}
          \and J. Maldonado\inst{7}
          \and L. Mancini\inst{11,3,12}
          \and D. Nardiello\inst{2,13}
          \and D. Sicilia\inst{6}
          \and A. Sozzetti\inst{3}
          \and A. Bignamini\inst{14}
          \and R. Cosentino\inst{15}
          \and C. Knapic\inst{14}
          \and A. F. Mart\'{i}nez Fiorenzano \inst{15}
          \and E. Molinari\inst{16}
          \and M. Pedani\inst{15}
          \and E. Poretti\inst{1,15}
}
   \institute{INAF - Osservatorio Astronomico di Brera, Via E. Bianchi, 46, 23807 Merate (LC), Italy 
   \and
   INAF - Osservatorio Astronomico di Padova, Vicolo dell'Osservatorio, 5, I-35122 Padova (PD), Italy 
   \and
   INAF - Osservatorio Astrofisico di Torino, Via Osservatorio 20, I-10025 Pino Torinese (TO), Italy 
   \and
   Department of Astronomy, University of Geneva, Chemin Pegasi 51, CH-1290 Versoix, Switzerland 
   \and
   INAF - Osservatorio Astronomico di Roma, Via Frascati 33, I-00078 Monte Porzio Catone (Roma), Italy 
   \and
   INAF - Osservatorio Astrofisico di Catania, Via S.Sofia 78, I-95123 Catania, Italy 
   \and
   INAF - Osservatorio Astronomico di Palermo, Piazza del Parlamento, 1, I-90134 Palermo, Italy 
   \and
   Instituto de Astrof\'{i}sica de Canarias (IAC), 38205 La Laguna, Tenerife, Spain  
   \and
   INAF - Osservatorio Astronomico di Capodimonte, via Moiariello 16, 80131 Napoli, Italy  
   \and
   Université Côte d'Azur, Observatoire de la Côte d'Azur, CNRS, Laboratoire Lagrange, Bd de l'Observatoire, CS34229, 06304 Nice Cedex 4, France 
   \and Department of Physics, University of Rome ``Tor Vergata'', Via della Ricerca Scientifica 1, I-00133, Rome, Italy 
   \and
    Max Planck Institute for Astronomy, K\"{o}nigstuhl 17, 69117, Heidelberg, Germany  
   \and
   Aix Marseille Univ, CNRS, CNES, LAM, Marseille, France 
   \and
   INAF – Osservatorio Astronomico di Trieste, via Tiepolo 11, 34143 Trieste, Italy 
   \and
   INAF - Fundaci\'{o}n Galileo Galilei, Rambla Jos\'{e} Ana Fernandez P\'{e}rez 7, 38712 Bre{\~n}a Baja (TF), Spain 
   \and
   INAF - Osservatorio Astronomico di Cagliari, via della Scienza 5, 09047 Selargius (CA), Italy 
   }

   \date{Received <date>; accepted <date>}

 
  \abstract
   {The leading spectrographs used for exoplanets' search and characterization offer online data reduction softwares (DRS) that yield as an ancillary result the full-width at half-maximum (FWHM) of the cross-correlation function (CCF) that is used to estimate the radial velocity of the host star. The FWHM also contains information on the stellar projected rotational velocity $v_{\mathrm{eq}}\sin{i_\star}$, if appropriately calibrated.}
   {We wanted to establish a simple relationship to derive the $v_{\mathrm{eq}}\sin{i_\star}$ directly from the FWHM computed by the HARPS-N DRS in the case of slow-rotating solar-like stars. This may also help to recover the stellar inclination $i_\star$, which in turn affects the exoplanets' parameters.}
   {We selected stars with an inclination of the spin axis compatible with 90 $\deg$ by looking at exoplanetary transiting systems with known small sky-projected obliquity: for these calibrators, we can presume that $v_{\mathrm{eq}}\sin{i_\star}$ is equal to stellar equatorial velocity $v_{\mathrm{eq}}$. We derived their rotational periods from photometric and spectroscopic time-series and their radii from SED fitting. This allowed us to recover their $v_{\mathrm{eq}}$, which we could compare to the FWHM values of the CCFs obtained both with G2 and K5 spectral type masks.}
   {We obtained an empirical relation for each mask: this can be used to derive $v_{\mathrm{eq}}\sin{i_\star}$ directly from FWHM values for slow rotators (FWHM < 20 km~s$^{-1}$). We applied our relations to 273 exoplanet-host stars observed with HARPS-N, obtaining homogeneous $v_{\mathrm{eq}}\sin{i_\star}$ measurements. When possible, we compared our results with the literature ones to confirm the reliability of our work. We were also able to recover or constrain $i_\star$ for 12 objects with no prior $v_{\mathrm{eq}}\sin{i_\star}$ estimation.}
   {We provide two simple empirical relations to directly convert the HARPS-N FWHM obtained with G2 and K5 mask to a $v_{\mathrm{eq}}\sin{i_\star}$ value. We tested our results on a statistically significant sample, and we found a good agreement with literature values found with more sophisticated methods  for stars with $\log g$ > 3.5. We also tried our relation on HARPS and SOPHIE data, and we conclude that it can be used as it is also on FWHM derived by HARPS DRS with G2 and K5 mask, and it may be adapted to the SOPHIE data as long as the spectra are taken in the high-resolution mode.}

   \keywords{Planetary systems --
             Techniques: spectroscopic --
             Stars: rotation  }
   \titlerunning{The GAPS programme at TNG XLV. Projected rotational velocities}
   \maketitle
%

\section{Introduction}\label{sec:intro}
Stable, high-resolution optical spectrographs are some of the leading instruments used for the search and characterization of the exoplanets: many of them are designed expressly for these studies (e.g. HARPS, HARPS-N, ESPRESSO), and as such they are equipped with dedicated data reduction softwares (DRS). One of the main deliverables of the DRS is the cross-correlation function (CCF) of the reduced spectra with a stellar mask chosen from the available library of spectral type templates (\citealp{Baranne1996,Pepe2002}).

The CCF allows computing the radial velocity of the host star with very high precision, and it also yields a number of additional parameters, such as the CCF's bisector span (which can be used as an activity indicator), the CCF's contrast, and the full-width at half-maximum (FWHM). The latter may be related to the stellar projected rotational velocity $v_{\mathrm{eq}}\sin{i_\star}$ if appropriately calibrated: in this paper, we present the work done to calibrate the FWHM of the CCF that is computed by the HARPS-N DRS \citep{Cosentino2014} using the G2 and K5 stellar masks.
HARPS-N is the high-resolution optical spectrograph installed at the Telescopio Nazionale Galileo (TNG) at the Roque de Los Muchachos Observatory (La Palma, Canary Islands, Spain).

The use of the CCF's FWHM to estimate the $v_{\mathrm{eq}}\sin{i_\star}$ is particularly important in the case of slowly rotating stars, for which the $v_{\mathrm{eq}}\sin{i_\star}$ computation via Fourier transform of the line profiles or fitting with a rotational profile is complicated by the combination of the rotational broadening with the effects of the resolution smearing ($\approx$ 2.6 km~s$^{-1}$ in the case of HARPS-N, R=115,000), and the micro- ($v_{\rm micro}$) and macro- ($v_{\rm macro}$) turbulence broadening. Slowly rotating solar-like and M-type stars are also among the main targets in the exoplanet field, therefore it is particularly important to have a reliable method to estimate the $v_{\mathrm{eq}}\sin{i_\star}$ for these objects in order to better characterize the host stars. Using the FWHM given by the HARPS-N DRS allows everyone to recover the $v_{\mathrm{eq}}\sin{i_\star}$ values directly for the HARPS-N archival data.

Once it is obtained, the $v_{\mathrm{eq}}\sin{i_\star}$ value may be used along with estimates of the stellar rotational period $P_{\mathrm{rot}}$ (for example from photometric time-series or spectroscopic time-series of activity indices) and the stellar radius $R_{\star}$ (derived for example from Spectral Energy Distribution (SED) fitting, see Sec. \ref{sec:calibrators}) to recover the stellar inclination $i_\star$, which heavily affects exoplanets' parameters \citep{Hirano2014}:
\begin{equation}
    i_\star = \mathrm{arcsin}(\frac{P_\mathrm{rot} \times  v_{\mathrm{eq}}\sin{i_\star}}{2\pi R_\star})
    \label{eq:inclination}
\end{equation}

The stellar inclination is a fundamental step also in computing the spin-orbit angle of exoplanetary systems, that is an important observational probe of the origin and evolution of the systems (e.g., \citealt{Queloz2000,Winn2005}).

The approach of exploiting known stellar radii and rotational periods to infer the rotational velocity and to calibrate the width of the CCF vs. $v_{\mathrm{eq}}\sin{i_\star}$ is not completely new, as it was previously adopted by \citet{Nordstrom2004}. However, in their case, the stellar inclination remains unknown and the additional uncertainty is treated statistically.
Instead, in our work we took advantage of the known viewing geometry of stars which host a transiting planet with an orbit inclination close to 90 $\deg$, and a good spin-orbit alignment as inferred by the measurement of the Rossiter-McLaughlin effect (\citealp{Rossiter1924,McLaughlin1924}). This allowed us to rely on a sample of objects for which the projected rotational velocity, linked to the CCF width, is similar to the equatorial velocity inferred from the rotational period and the stellar radius. Furthermore, the selection of a sample of transiting planets ensures the availability of high-quality photometric data (which were taken for the planet search itself) and in most cases of additional relevant literature studies from follow-up observations.

This paper is organized as follows: in Section~\ref{sec:calibrators} we describe the selection procedure for our calibrators. We used them in Section~\ref{sec:relation} to create our empirical relation, and then we applied it to a large set of exoplanet host stars in Section~\ref{sec:vsini}. We test the applicability of our relation to other spectrographs in Section~\ref{sec:other_spectrographs}, and finally we present our conclusions in Section~\ref{sec:conclusions}.

\section{Calibrators' selection and characterization}\label{sec:calibrators}
To calibrate our empirical relation as accurately as possible, we relied on a very strict selection of calibrators. We queried the NASA exoplanet archive \footnote{\url{https://exoplanetarchive.ipac.caltech.edu/}} to obtain a list of all known exoplanet host stars with \textit{a)} declination $> -25\deg$ (to ensure they were observable with the TNG), and \textit{b)} an absolute value of the system sky-projected obliquity $\lambda$ smaller than 30 $\deg$, as derived from the Rossiter-McLaughlin effect and reported in the TEPCat catalog \citep{southworth2011}. The latter value is a compromise between the need to have systems that can be considered aligned in such a way that the stellar projected rotational velocity $v_{\mathrm{eq}}\sin{i_\star}$ can be considered approximately equal to the stellar equatorial velocity $v_{\mathrm{eq}}$, and the need to have a good number of useful calibrators (at least some tens of objects).

This selection resulted in a list of 66 targets. We then searched the TNG archive for public HARPS-N spectra of these stars, to combine with the proprietary data obtained within the Global Architecture of Planetary Systems (GAPS) program, which is an Italian project dedicated to the search and characterization of exoplanets (PI G. Micela; \citealt{Covino2013}). We thus found 44 stars with useful HARPS-N CCFs.

The stellar masks available in the DRS library are optimized for main sequence stars with stellar types G2, K5, and M2. With the new upgrades to the DRS, more masks are starting to be available for different spectral types, and they will have to be calibrated accordingly, but in this work we focused on the original masks that have been used so far, and that are still available in the DRS.
Unfortunately, the M2 CCFs are useless for our purposes because the use of the M2 mask results in deformed CCF profiles with large bumps in the wings. In a previous work \citep{Rainer2020}, we created an improved M-type mask to overcome this problem, but we will not consider this mask here because it is not publicly available: our scope is to enable astronomers to use the public HARPS-N archival data. Thus, we focused on the G2, and K5 CCFs: while this optimized our work for solar-like stars, still some M-type stars may be reduced using the K5 mask in order to recover the $v_{\mathrm{eq}}\sin{i_\star}$ estimate from the CCF FWHM.

Our selection criteria ensure that $sin{i_\star} \approx 1$, which means that we can consider $v_{\mathrm{eq}}\sin{i_\star}$ $\approx$ $v_{\mathrm{eq}}$ for all our calibrators. If we are able to estimate the equatorial velocity $v_{\mathrm{eq}}$, then we can build a relation between FWHM and $v_{\mathrm{eq}}\sin{i_\star}$ in a straightforward way. In order to compute $v_{\mathrm{eq}}$ we needed estimates of the rotational periods $P_{\mathrm{rot}}$ and the radii $R_{\star}$ of our calibrators:
\begin{equation}
    \label{eq:veq}
    v_\mathrm{eq}=\frac{2\pi \times R_{\star}}{P_\mathrm{rot}} \, .
\end{equation}

We derived the rotation period $P_{\mathrm{rot}}$ mainly from TESS \citep{Ricker2015} and SuperWASP \citep{Butters2010} photometry. In the case of TESS, we used the PDCSAP light curves \citep{Stumpe2012} as downloaded from the MAST archive\footnote{\url{ https://mast.stsci.edu/portal/Mashup/Clients/Mast/Portal.html}}, where systematic artifacts are likely removed by the PDCSAP pipeline. PDCSAP light curves were analysed using the Generalized Lomb-Scargle periodogram (GLS; \citep{Zechmeister2009} and the detected periods are listed in  Table\,\ref{tab:calibrators}
In the case of the SuperWASP photometric time-series, we first disregarded possible outliers, i.e. data points that deviated more than 3 standard deviations from the mean of the whole data series. Then, we computed a filtered version of the light curve by means of a sliding median boxcar filter with a boxcar extension equal to 2 hours. This filtered light curve was then subtracted from the original light curve, and all the points deviating more than 3 standard deviations of the residuals were discarded. Finally, we computed normal points by binning the data on time intervals having the duration of about 2 hours.
The rotation period search was performed by using the GLS and the CLEAN \citep{Roberts87} periodogram analysis. All the periodicities detected by GLS, with a false alarm probability smaller than 0.1\%  \citep[see][]{Horne86}, and recovered with the same value, within the uncertainty, also by CLEAN, were considered as the star's rotation period and listed in Table\,\ref{tab:calibrators}. To compute the error associated with the period, we followed the method used by \citet{Lamm2004}.
\begin{table*}
    \caption{Calibrators' list. HAT-P-2 is present here, but not used as a calibrator because of its large FWHM value (> 20 km~s$^{-1}$).}
    \label{tab:calibrators}
    \centering
    \begin{tabular}{c c c c c c c c c}
    \hline
    \hline 
    Name & $\lambda$ & $T_{\mathrm{eff}}$ & $\log g$ & [Fe/H] &$P_{\mathrm{rot}}$ & $R_{\star}$ & $v_{\rm micro}$ & $v_{\rm macro}$ \\
    & [deg] & [K] & [dex] & [dex] & [days] & [$R_{\sun}$] & [km~s$^{-1}$] & [km~s$^{-1}$] \\
    \hline
    HAT-P-1 & 3.7 & 5980$\pm$49$^4$ & 4.36$\pm$0.01$^4$ & 0.13$\pm$0.008$^4$ & 48$\pm$5$^1$ & 1.273$\pm$0.065$^3$ & 1.19$^3$ & 3.92$^3$\\
    HAT-P-2 & 9.0 & 6380$\pm$0$^5$ & 4.16$\pm$0.02$^6$ & 0.13$\pm$0.008$^7$ & 97$\pm$10$^1$ & 1.684$\pm$0.029$^3$ & 1.68$^3$ & 5.90$^3$\\
    & & & & & 2.82$\pm$0.05$^2$ & & & \\
    & & & & & 96$\pm$15$^3$ & & & \\
    HAT-P-3 & 21.2 & 5185$\pm$80$^6$ & 4.56$\pm$0.03$^6$ & 0.24$\pm$0.08$^8$ & 28$\pm$2$^1$ & 0.861$\pm$0.015$^3$ & 0.59$^3$ & 2.02$^3$\\
    & & & & & 40$\pm$2$^1$ & & & \\
    HAT-P-8 & -17.0 & 6200$\pm$80$^6$ & 4.15$\pm$0.03$^6$ & 0.01$\pm$0.08$^7$ & 4.25$\pm$0.05$^1$ & 1.546$\pm$0.027$^3$ & 1.48$^3$ & 5.13$^3$\\
    HAT-P-13 & 1.9 & 5653$\pm$90$^7$ & 4.13$\pm$0.04$^9$ & 0.41$\pm$0.08$^7$ & 30$\pm$3$^1$ & 1.824$\pm$0.038$^3$ & 0.99$^3$ & 3.57$^3$ \\
    HAT-P-16 & -2.0 & 6158$\pm$80$^{10}$ & 4.34$\pm$0.03$^{10}$ & 0.17$\pm$0.08$^{10}$ & 12.7$\pm$0.5$^1$ & 1.221$\pm$0.019$^3$ & 1.37$^3$ & 4.58$^3$ \\
    HAT-P-17 & 19.0 & 5246$\pm$80$^6$ & 4.53$\pm$0.02$^6$ & 0.0$\pm$0.08$^7$ & 33$\pm$5$^1$ & 0.87$\pm$0.018$^3$ & 0.62$^3$ & 2.35$^3$\\
    & & & & & 25$\pm$8.3$^2$ & & & \\
    HAT-P-20 & -8.0 & 4595$\pm$80$^{11}$ & 4.63$\pm$0.02$^{11}$ & 0.22$\pm$0.09$^{12}$ & 14.48$\pm$0.02$^{12}$ & 0.722$\pm$0.011$^3$ & 0.45$^3$ & 1.53$^3$\\
    & & & & & 14.44$\pm$0.07$^1$ & & & \\
    HAT-P-22 & -2.1 & 5302$\pm$80$^6$ & 4.36$\pm$0.04$^6$ & 0.30$\pm$0.09$^8$ & 28.7$\pm$0.04$^8$ & 1.075$\pm$0.024$^3$ & 0.70$^3$ & 2.71$^3$\\
    & & & & & 37$\pm$1$^1$ & & & \\
    HD 17156 & 10.0 & 6040$\pm$24$^6$ & 4.20$\pm$0.06$^6$ & 0.24$\pm$0.03$^{13}$ & 12.8$\pm$0.0$^{13}$ & 1.52$\pm$0.033$^3$ & 1.30$^3$ & 4.44$^3$ \\
    HD 63433 & 8.0 & 5640$\pm$74$^{14}$ & 4.53$\pm$0.09$^{14}$ & 0.017$\pm$0.017$^{15}$ & 6.45$\pm$0.05$^{14}$ & 0.911$\pm$0.021$^3$ & 0.85$^3$ & 2.74$^3$ \\
    & & & & & 6.25$\pm$0.93$^2$ & & & \\
    HD 189733 & -0.31 & 5052$\pm$16$^6$ & 4.49$\pm$0.05$^6$ & 0.03$\pm$0.08$^7$ & 11.95$\pm$0.01$^7$ & 0.787$\pm$0.036$^3$ & 0.56$^3$ & 1.86$^3$\\
    HD 209458 & 1.58 & 6091$\pm$10$^6$ & 4.45$\pm$0.02$^6$ & 0.0$\pm$0.05$^7$ & 10.65$\pm$0.75$^7$ & 1.178$\pm$0.028$^3$ & 1.27$^3$ & 4.11$^3$ \\
    K2-29 & 1.5 & 5358$\pm$38$^{16}$ & 4.54$\pm$0.01$^{16}$ & 0.03$\pm$0.05$^{16}$ & 10.79$\pm$0.02$^{16}$ & 0.847$\pm$0.019$^3$ & 0.67$^3$ & 2.38$^3$\\
    & & & & & 10.41$\pm$0.07$^1$ & & & \\
    K2-34 & -1.0 & 6071$\pm$90$^{17}$ & 4.18$\pm$0.02$^{17}$ & -- &  7.9$\pm$0.2$^1$ & 1.43$\pm$0.023$^3$ & 1.33$^3$ & 4.58$^3$ \\
    Kepler-25 & 9.4 & 6354$\pm$27$^{18}$ & 4.29$\pm$0.01$^{18}$ & 0.11$\pm$0.03$^{18}$ & 23.147$\pm$0.039$^{19}$ & 1.737$\pm$0.1$^3$ & 1.61$^3$ & 5.52$^3$\\
    Qatar-1 & -8.4 & 5013$\pm$93$^{20}$ & 4.55$\pm$0.01$^{20}$ & 0.2$\pm$0.1$^7$ & 23.7$\pm$0.1$^{20}$ & 0.792$\pm$0.013$^3$ & 0.53$^3$ & 1.82$^3$\\
    Qatar-2 & 15.0 & 4645$\pm$50$^{21}$ & 4.53$\pm$0.01$^{21}$ & 0.02$\pm$0.08$^{21}$ & 18.0$\pm$0.2$^{22}$ & 0.721$\pm$0.012$^3$ & 0.48$^3$ & 1.56$^3$\\
    TrES-4 & 6.3 & 6200$\pm$75$^6$ & 4.06$\pm$0.02$^6$ & 0.28$\pm$0.09$^7$ & 26.2$\pm$2$^1$ & 1.984$\pm$0.028$^3$ & 1.51$^3$ & 5.31$^3$\\
    WASP-11 & 7.0 & 4800$\pm$100$^6$ & 4.45$\pm$0.02$^6$ & 0.12$\pm$0.09$^7$ & 15.26$\pm$0.07$^1$ & 0.857$\pm$0.018$^3$ & 0.52$^3$ & 1.64$^3$\\
    WASP-13 & 8.0 & 5950$\pm$70$^6$ & 4.06$\pm$0.01$^6$ & 0.0$\pm$0.2$^7$ & 9.66$\pm$0.9$^1$ & 1.581$\pm$0.024$^3$ & 1.25$^3$ & 4.43$^3$\\
    WASP-14 & -14.0 & 6475$\pm$100$^6$ & 4.07$\pm$0.02$^6$ & 0.0$\pm$0.2$^7$ & 22$\pm$3$^1$ & 0.983$\pm$0.037$^3$ & 1.83$^3$ & 6.55$^3$\\
    WASP-32 & -2.0 & 6140$\pm$95$^6$ & 4.40$\pm$0.02$^6$ & 0.13$\pm$0.1$^7$ & 11.6$\pm$1.0$^7$ & 1.01$\pm$0.077$^3$ & 1.33$^3$ & 4.39$^3$\\
    WASP-43 & 3.5 & 4400$\pm$200$^{23}$ & 4.49$\pm$0.13$^6$ & 0.05$\pm$0.17$^7$ & 15.6$\pm$0.4$^7$ & 0.679$\pm$0.014$^3$ & 0.50$^3$ & 1.46$^3$\\
    & & & & & 13.3$\pm$5.1$^2$ & & & \\
    WASP-69 & 0.4 & 4700$\pm$50$^6$ & 4.50$\pm$0.15$^6$ & 0.15$\pm$0.08$^7$ & 23.07$\pm$0.16$^7$ & 0.836$\pm$0.014$^3$ & 0.49$^3$ & 1.58$^3$\\
    WASP-84 & -0.3 & 5314$\pm$88$^{24}$ & 4.40$\pm$0.13$^{24}$ & 0.0$\pm$0.1$^7$ & 14.36$\pm$0.35$^7$ & 0.822$\pm$0.011$^3$ & 0.69$^3$ & 2.63$^3$\\
    XO-2N & 7.0 & 5340$\pm$50$^{25}$ & 4.43$\pm$0.01$^{26}$ & 0.43$\pm$0.05$^7$ & 28.6$\pm$1.3$^7$ & 0.998$\pm$0.014$^3$ & 0.70$^3$ & 2.59$^3$\\
    & & & & & $35\pm3^1$  & & & \\
    \hline
    \end{tabular}
    \tablebib{$^1$~SuperWASP;$^2$~TESS;$^3$~this work;$^4$~\cite{Nikolov2014};$^5$~\cite{Ment2018};$^6$~\cite{Stassun2017};$^7$~\cite{Bonomo2017};$^8$~\cite{Mancini2018};$^9$~\cite{Sada2016};$^{10}$~\cite{Buchhave2010};$^{11}$~\cite{Bakos2011};$^{12}$~\cite{Esposito2017};$^{13}$~\cite{Fischer2007};$^{14}$~\cite{Mann2020};$^{15}$~\url{https://exofop.ipac.caltech.edu};$^{16}$~\cite{Santerne2016};$^{17}$~\cite{Livingston2018};$^{18}$~\cite{Benomar2014};$^{19}$~\cite{McQuillan2013};$^{20}$~\cite{Collins2017};$^{21}$~\cite{Mocnik2017};$^{22}$~\cite{Mancini2014};$^{23}$~\cite{Hellier2011};$^{24}$~\cite{Anderson2014};$^{25}$~\cite{Southworth2012};$^{26}$~\cite{Crouzet2012}}
\end{table*}

We also checked the spectroscopic activity indicators time-series: we investigated the $R'_{\rm HK}$ activity index using GLS. In general, we did not find any conclusive results given that for most stars only a small number of observations sparsely obtained over a few years were available. In a few cases, the periodogram analysis provided $P_{\mathrm{rot}}$ detection, which was always consistent with the photometrically determined period. For the sake of sample homogeneity, we thus considered only the photometric periods.

\begin{figure}
    \centering
    \includegraphics[width=\columnwidth]{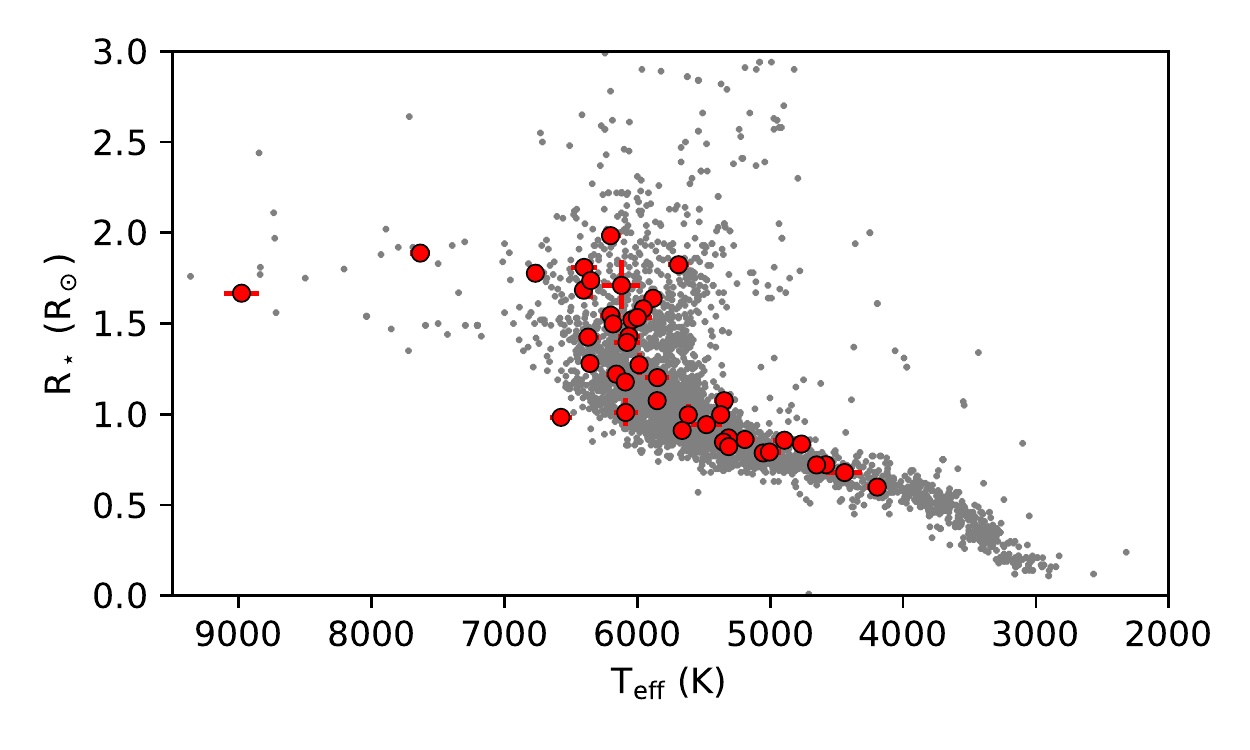}
    \caption{Comparison in the $T_{\mathrm{eff}}$-$R_{\star}$ parameter space between the sample of stars analysed in this work (red circles) and the currently known exoplanet-host stars (grey dots) as retrieved from the NASA Exoplanet Catalog.}
    \label{fig:sample-HR}
\end{figure}

\begin{figure}
    \centering
    \includegraphics[width=\columnwidth]{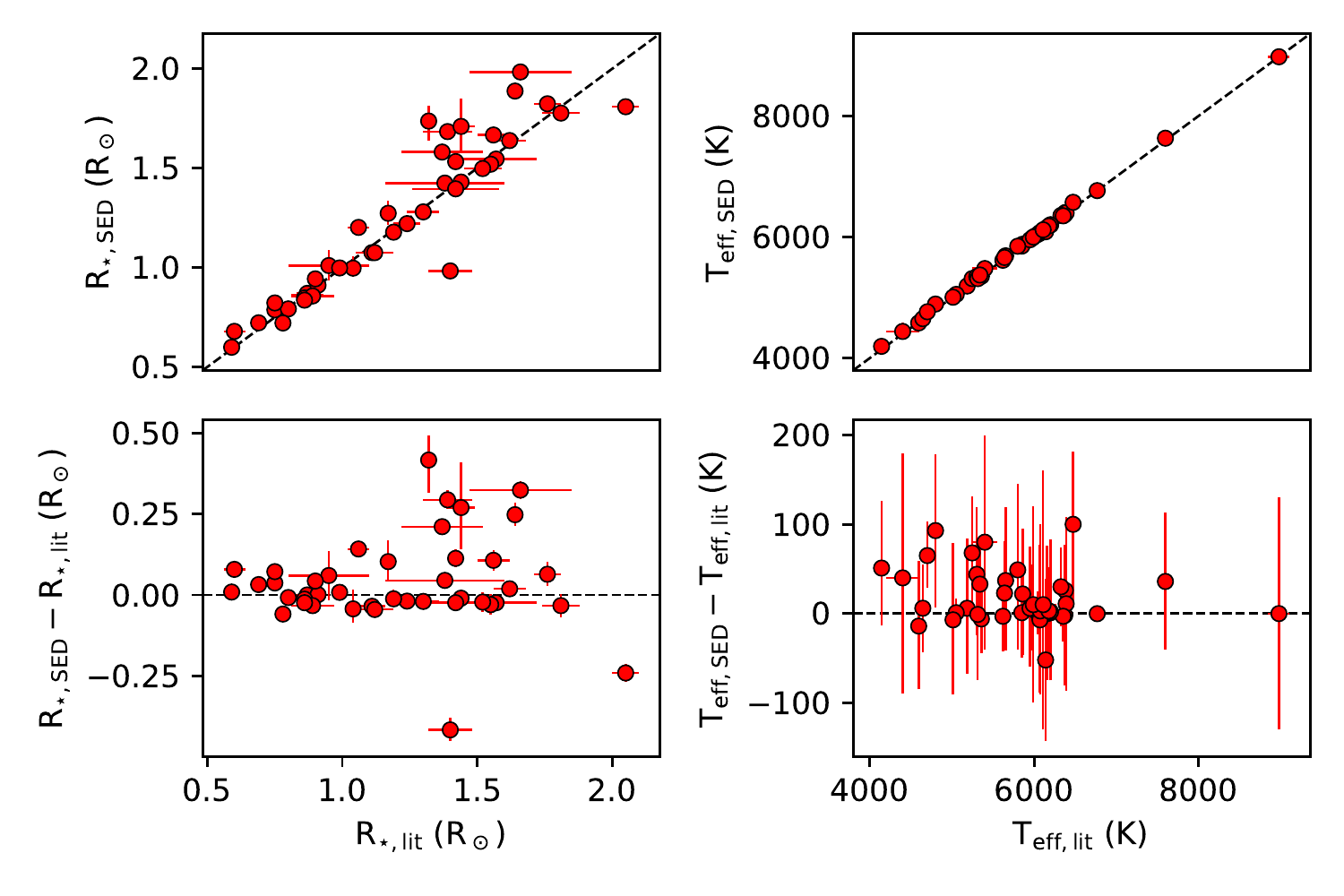}
    \caption{Comparison of the stellar radii and effective temperatures obtained via the SED fitting described in Sect.~\ref{sec:calibrators} with literature values. \textit{Upper panels:} correlations plot between our values and literature ones for $R_{\star}$ (left panel) and $T_{\mathrm{eff}}$ (right panel) \textit{Lower panels:} residual plots showing the difference between our values and literature ones.}
    \label{fig:radius-teff-comparison}
\end{figure}

The stellar radii $R_{\star}$ were obtained by fitting the Spectral Energy Distribution (SED) via the MESA Isochrones and Stellar Tracks \citep[MIST,][]{Dotter2016,Choi2016} through the \texttt{EXOFASTv2} suite \citep{Eastman2019}. Specifically, we fitted the available archival magnitudes of each star in the sample imposing Gaussian priors on the effective temperature $T_{\mathrm{eff}}$ and metallicity [Fe/H] based on the respective literature values listed in Table~\ref{tab:calibrators} and on the parallax $\pi$ based on the Gaia~EDR3 astrometric measurement \citep{Gaia2016,Gaia2021}. Since the SED primarily constrains $R_{\star}$ and $T_{\mathrm{eff}}$, the stellar parameters are simultaneously constrained by the SED and the MIST isochrones, and a penalty for straying from the MIST evolutionary tracks ensures that the resulting star realization is physical in nature \citep[see][for more details on the method]{Eastman2019}. In Fig.~\ref{fig:sample-HR} we show our results compared with the $R_{\star}$ and $T_{\mathrm{eff}}$ of the exoplanet-host stars present in the NASA archive, while in Fig.~\ref{fig:radius-teff-comparison} we show the correlation and residuals between our values and those from the literature.

We also checked the literature for asteroseismic and interferometric radii, which we found for HD 17156 \citep[asteroseismic $R_{\star}$ = 1.5007$\pm$0.0076 $R_{\sun}$,][]{Nutzman_2010}, HD 189733 and HD 209458 \citep[interferometric $R_{\star}$ 0.805$\pm$0.016 and 1.203$\pm$0.06 $R_{\sun}$ respectively,][]{Boyajian2014}: they are in good agreement with our results.

Thus we obtained our semi-final calibrators' list, which is shown in Table~\ref{tab:calibrators}: 27 stars with known $P_{\mathrm{rot}}$ and $R_{\star}$. In the end, all our calibrators have $\lambda < 21.2$ degrees, strengthening our assumption of  $v_{\mathrm{eq}} \approx v_{\mathrm{eq}}\sin{i_\star}$.

We also checked the Gaia DR3 archive to ensure that we are working with single stars: K2-29 has a fainter companion separated by $\approx$4.4 arcsec with $\Delta$V=1.8, and TrES-4 has a fainter companion separated by $\approx$1.6 arcsec with $\Delta$V=4.9. We considered that in both cases the combination of the faintness and the distance of the companions allowed us to keep the stars in our calibrators' list.

Using the stellar parameters $T_{\mathrm{eff}}$ and $\log g$ from the literature, we estimated the micro- ($v_{\rm micro}$) and macro- ($v_{\rm macro}$) turbulence velocities for each object. In particular, $v_{\rm micro}$ was obtained with \cite{Adibekyan2012} relationships valid for stars with 4500 < $T_{\mathrm{eff}}$ < 6500\ K, 3.0 < $\log g$ < 5.0, and -1.4 < [Fe/H] < 0.5 dex. Regarding $v_{\rm macro}$, it was computed with the calibration obtained by \cite{Doyle2014} using asteroseismic rotational velocities for the stars with $T_{\mathrm{eff}}$ > 5700 K, while for the stars with $T_{\mathrm{eff}}$ < 5700 K we used the empirical relationship by \cite{Breweretal2016}. Both relations are valid for dwarf stars \citep[see also][]{Biazzo2022}. 
To estimate the errors on our $v_{\rm micro}$ and $v_{\rm macro}$, we considered the root-mean-square error (rms) given in the papers, which is larger than the errors derived from the parameters. The rms are 0.18 km~s$^{-1}$ for $v_{\rm micro}$, 0.73 km~s$^{-1}$ for $v_{\rm macro}$ from \cite{Doyle2014} ($T_{\mathrm{eff}}$ > 5700 K), and 0.5 km~s$^{-1}$ for $v_{\rm macro}$ from \cite{Breweretal2016} ($T_{\mathrm{eff}}$ < 5700 K).

We note that HAT-P-2 has $P_{\mathrm{rot}}$=2.82$\pm$0.05 days from TESS photometry, but a completely different value from SuperWASP (97$\pm$10 days). Applying Eq.~\ref{eq:veq}, the TESS value yields $v_{\mathrm{eq}}$=30.12 km~s$^{-1}$, and the SuperWASP value $v_{\mathrm{eq}}$=0.88 km~s$^{-1}$. The TESS value is nearer to the $v_{\mathrm{eq}}\sin{i_\star}$ = 20.12$\pm$0.9 km~s$^{-1}$ result obtained from the Fourier transform of the CCF and with the 20.8$\pm$0.03 km~s$^{-1}$ value from the literature \citep{Bonomo2017}, but there is still a large discrepancy. In any case, this fast rotation excludes this star from being a useful calibrator (see Section~\ref{sec:relation}): the final calibrators' list thus contains the stars in Table~\ref{tab:calibrators} with the exception of HAT-P-2.


\section{Creating the empirical relation}\label{sec:relation}
In order to create our empirical relation, we used the following inputs:
\begin{itemize}
    \item the FWHM of the CCFs of the HARPS-N spectra, as computed by the HARPS-N DRS and stored in the keyword \texttt{HIERARCH TNG DRS CCF FWHM} of the CCF FITS files;
    \item the stellar radii $R_{\star}$ from Table~\ref{tab:calibrators};
    \item the rotational periods $P_{\mathrm{rot}}$ from Table~\ref{tab:calibrators};
    \item $v_{\rm micro}$ and $v_{\rm macro}$ from Table~\ref{tab:calibrators}.
\end{itemize}

Using the archival CCFs, we are limited by the standard CCF half-window of the HARPS-N DRS (20 km~s$^{-1}$): while it may be manually changed, the majority of the archival data will have this value. We also note that a more precise $v_{\mathrm{eq}}\sin{i_\star}$ could be recovered for faster rotating stars using rotational fitting or the Fourier transform method, instead of any empirical relation. We thus limit the applicability range of our relation to FWHM up to 20 km~s$^{-1}$, which is a slightly larger value than the maximum FWHM that can be reliably computed with an half-window of 20 km~s$^{-1}$, i.e. $\approx$16-18 km~s$^{-1}$.

To check this applicability range, we built a range of synthetic CCF profiles by convolving a Gaussian function with the same FWHM of the HARPS-N resolution ($\approx$2.6 km~s$^{-1}$) with different rotational profiles ($v_{\mathrm{eq}}\sin{i_\star}$ ranging from 0.2 to 50 km~s$^{-1}$ with a step of 0.2 km~s$^{-1}$). The rotational profiles were built using the following equation from \cite{Gray2008}:

\begin{equation}
    f(x) = 1-a \frac{2(1-u) \sqrt{1-\left( \frac{x-x_0}{x_l}\right) ^{2}} + 0.5\pi u \left[1- \left(\frac{x-x_0}{x_l}\right)^{2}\right]} {{\pi}x_l \left(1-\frac{u}{3} \right)},
    \label{eq:gray}
\end{equation}
where $a$ is the depth of the profile, $x_0$ the centre (i.e., the RV value), $x_l$ the $v_{\mathrm{eq}}\sin{i_\star}$ of the star, $u$ the linear limb darkening (LD) coefficient, which we kept fixed as $u$=0.6.

\begin{figure*}
\includegraphics[width=\textwidth]{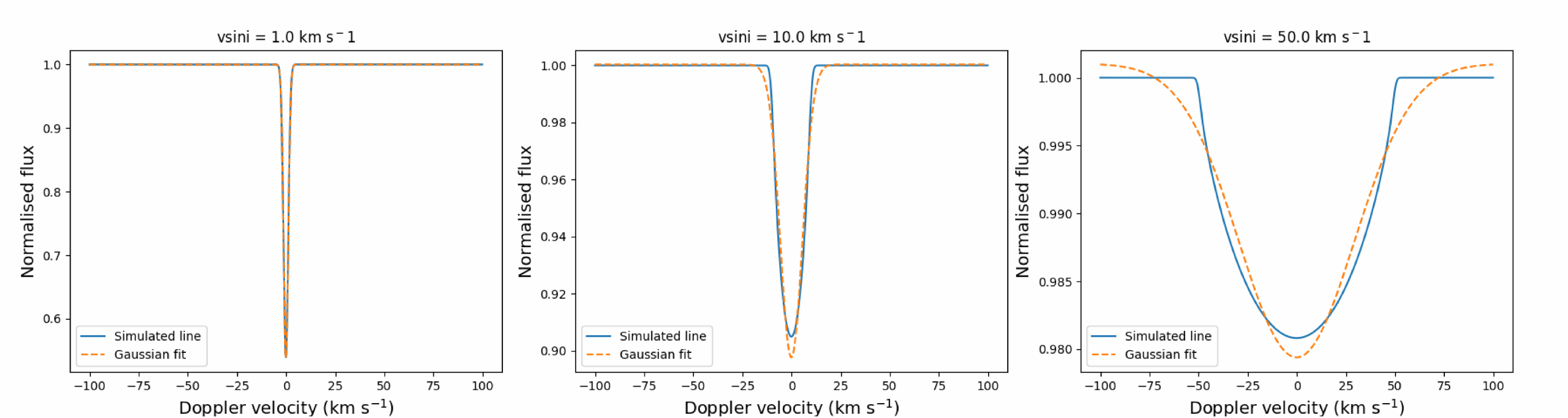}
\caption{Simulated CCF (solid blue line) and Gaussian fitting (dashed orange line). \textit{Left:} input value $v_{\mathrm{eq}}\sin{i_\star}$ = 1 km~s$^{-1}$. \textit{Center:} input value $v_{\mathrm{eq}}\sin{i_\star}$ = 10 km~s$^{-1}$. \textit{Right:} input value $v_{\mathrm{eq}}\sin{i_\star}$ = 50 km~s$^{-1}$.}
\label{fig:simulation}
\end{figure*}

We fitted the resulting profiles with a Gaussian (see Fig.~\ref{fig:simulation}) and compared the Gaussian FWHM with the input $v_{\mathrm{eq}}\sin{i_\star}$ to check their correlation. We chose a Gaussian fit to be consistent with HARPS-N DRS, that recovers both the radial velocity and the FWHM with a Gaussian fit of the CCF.

\begin{figure}
    \centering
    \includegraphics[width=\columnwidth]{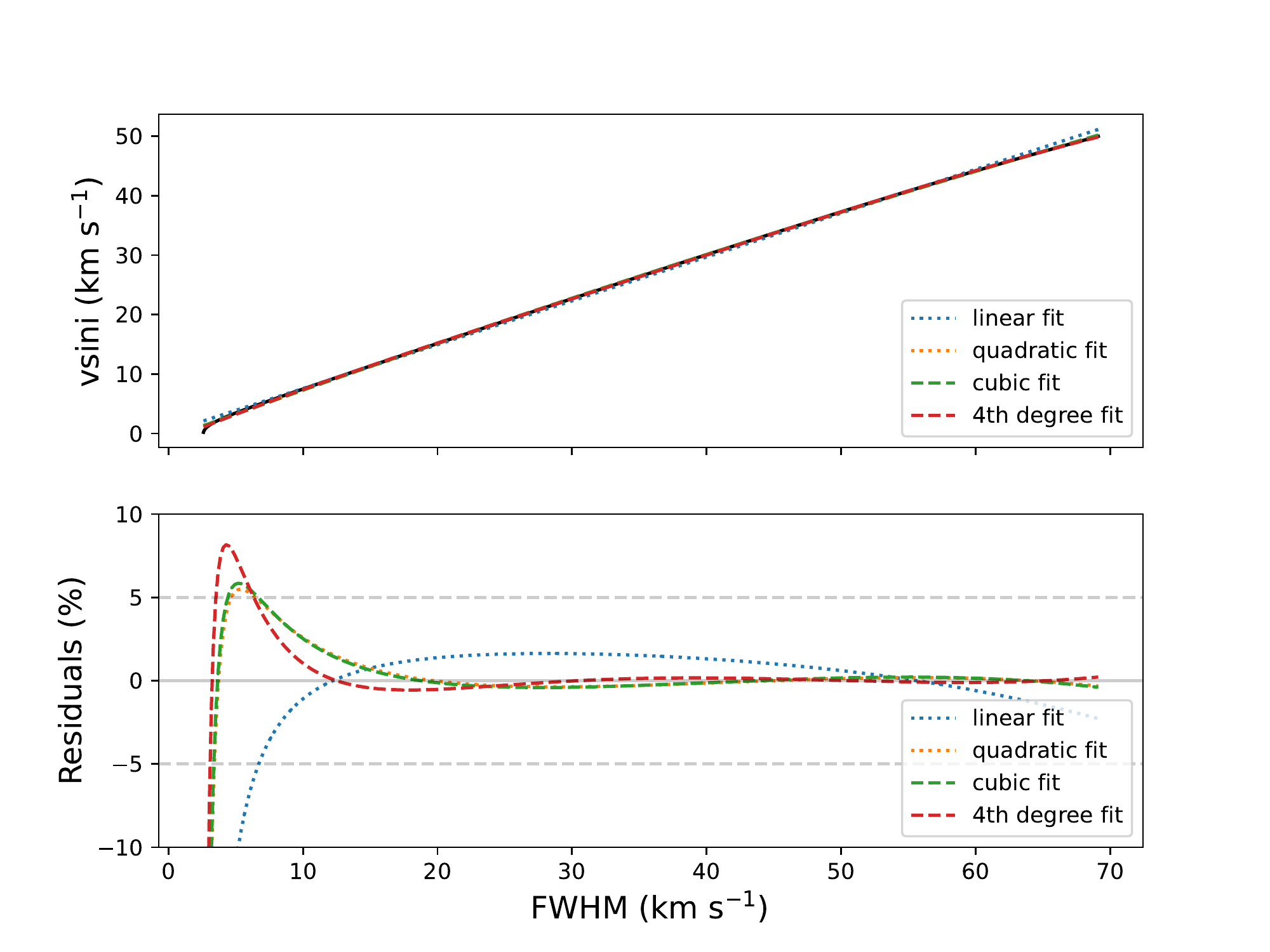}
    \caption{Correlation between the Gaussian fit's FWHM and the input $v_{\mathrm{eq}}\sin{i_\star}$ of the synthetic line profiles in the whole 0-50 km~s$^{-1}$ $v_{\mathrm{eq}}\sin{i_\star}$ (0-70 km~s$^{-1}$ FWHM) range. \textit{Upper panel:} correlation between FWHM and $v_{\mathrm{eq}}$ (black line) and relative linear fit (blue dotted line), quadratic fit (orange dotted line), cubic fit (green dashed line) and 4$^{\mathrm{th}}$ degree polynomial fit (red dashed line). \textit{Lower panel:} residuals of the fits. The horizontal grey lines outline the 5\% difference between the fit and the data}
    \label{fig:simulated_fit}
\end{figure}

\begin{figure}
    \centering
    \includegraphics[width=\columnwidth]{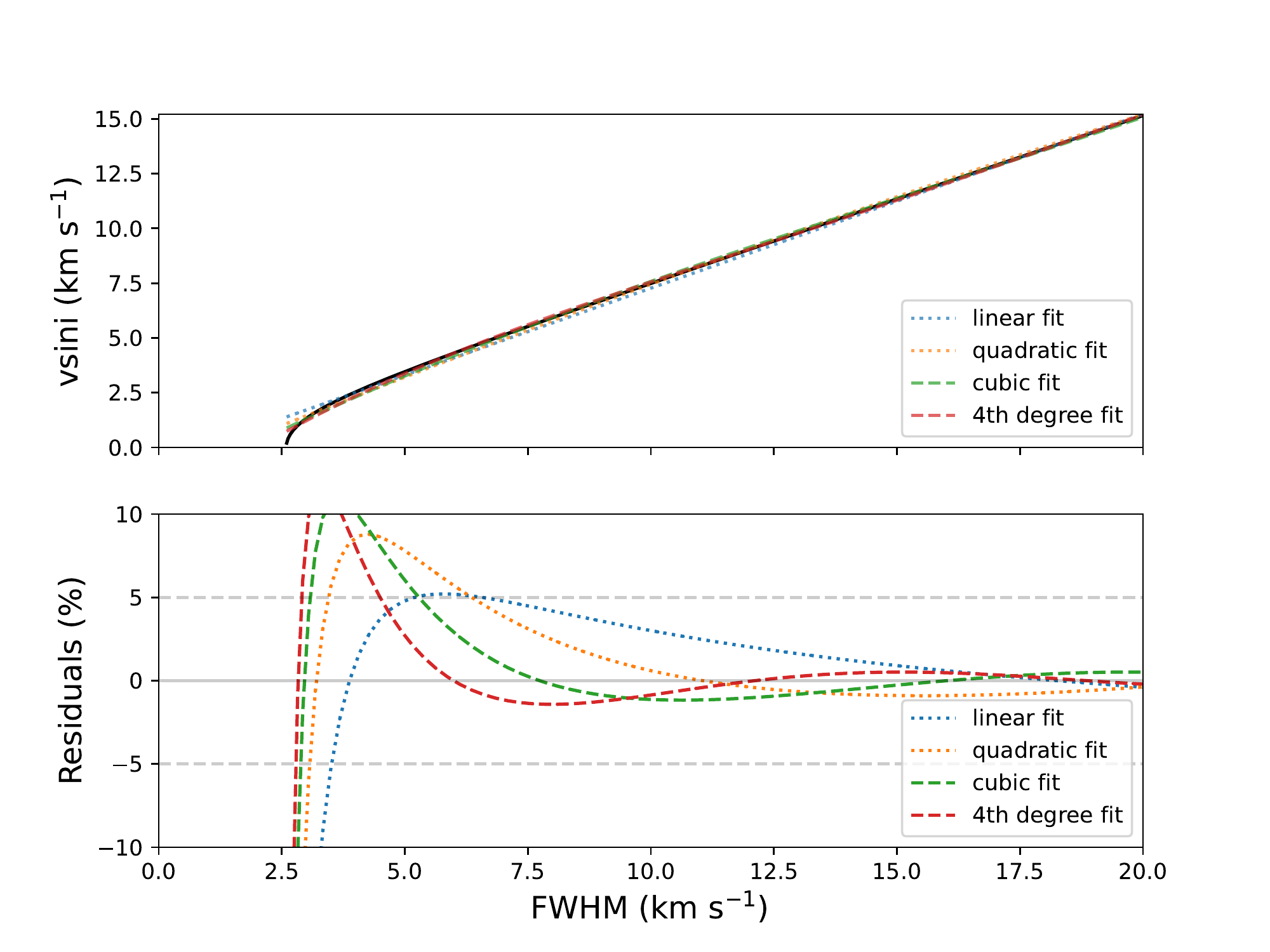}
    \caption{Same as Fig. \ref{fig:simulated_fit}, but limited to the 0-20 km~s$^{-1}$ FWHM range.}
    \label{fig:simulated_fit_zoom}
\end{figure}

Using a single fit for the whole range resulted in some discrepancy at the borders, in particular for low FWHM values (FWHM < 6.5 km~s$^{-1}$), i.e. the range we are more interested in (see Fig.~\ref{fig:simulated_fit}). As such we decided to try and improve the fit at lower values and limit our FWHM fitting range to 0-20 km~s$^{-1}$: in this case, while higher-order polynomials behave well enough down to FWHM=5km~s$^{-1}$, the linear fit residuals lie below 5\% down to FWHM=3.5km~s$^{-1}$ (see Fig.~\ref{fig:simulated_fit_zoom}). Considering that we have a small sample of calibrators (which hinders our ability to constrain an high degree polynomial), and that the linear fit recovers the $v_{\mathrm{eq}}\sin{i_\star}$ values with a 5\% error at worst, we can then reasonably assume that using a linear fit on the calibrators with FWHM < 20 km~s$^{-1}$ would give us useful results.

This is a simple test, that does not take into account all the other non-constant cause of broadening: for example the effects of $v_{\rm micro}$ and $v_{\rm macro}$, that highly depend on the stellar type, are not considered. A more detailed test would involve studying the CCFs obtained on a range of synthetic spectra with different $v_{\mathrm{eq}}\sin{i_\star}$ and stellar parameters: unfortunately the HARPS-N DRS works only on real raw HARPS-N data, so we cannot perform this analysis. However, we were still able to test our final results in this sense, because while our calibrators' sample is quite small, the total number of stars for which we computed $v_{\mathrm{eq}}\sin{i_\star}$, and that have literature values of $v_{\mathrm{eq}}\sin{i_\star}$ to compare to, is large enough to allow us to look for trends or misbehaviour related to the stellar parameters (see Sec.~\ref{sec:vsini}).

Taking into account all the previous considerations, such as the default half-window value of the CCFs, the aim to optimize the FWHM-$v_{\mathrm{eq}}\sin{i_\star}$ relation for the lower FWHM values, and above all the small sample of calibrators of which only one object (HAT-P-2) has FWHM > 20km~s$^{-1}$, we then excluded HAT-P-2 from the final calibrators' list and consider our work reliably applicable only for FWHM < 20 km~s$^{-1}$.

We created our relation first by using the CCFs computed with the G2 mask, and then we repeated the work described hereafter also for the K5 CCFs. We built four data sets:
\begin{itemize}
    \item the original FWHM computed by the DRS (FWHM$_\mathrm{DRS}$),
    \item the FWHM$_\mathrm{DRS}$ minus the $v_{\rm micro}$ broadening:
\begin{equation}
    \mathrm{FWHM}_\mathrm{mic} = \sqrt{{{\mathrm{FWHM}}_\mathrm{DRS}}^2 - \nu_m^2}
\end{equation}
\item the FWHM$_\mathrm{DRS}$ minus the $v_{\rm macro}$ broadening:
\begin{equation}
    \mathrm{FWHM}_\mathrm{mac} = \sqrt{{{\mathrm{FWHM}}_\mathrm{DRS}}^2 - \nu_M^2}
\end{equation}
\item and the FWHM$_\mathrm{DRS}$ minus both $v_{\rm micro}$ and $v_{\rm macro}$ broadening:
\begin{equation}
    \mathrm{FWHM}_\mathrm{mic+mac} = \sqrt{{{\mathrm{FWHM}}_\mathrm{DRS}}^2 - \nu_m^2 - \nu_M^2}
\end{equation}
\end{itemize}
 We considered also removing the instrumental broadening, but since this is a constant effect in HARPS-N spectra it will simply be included in the empirical relation.

\begin{figure*}
    \centering
    \includegraphics[width=\textwidth]{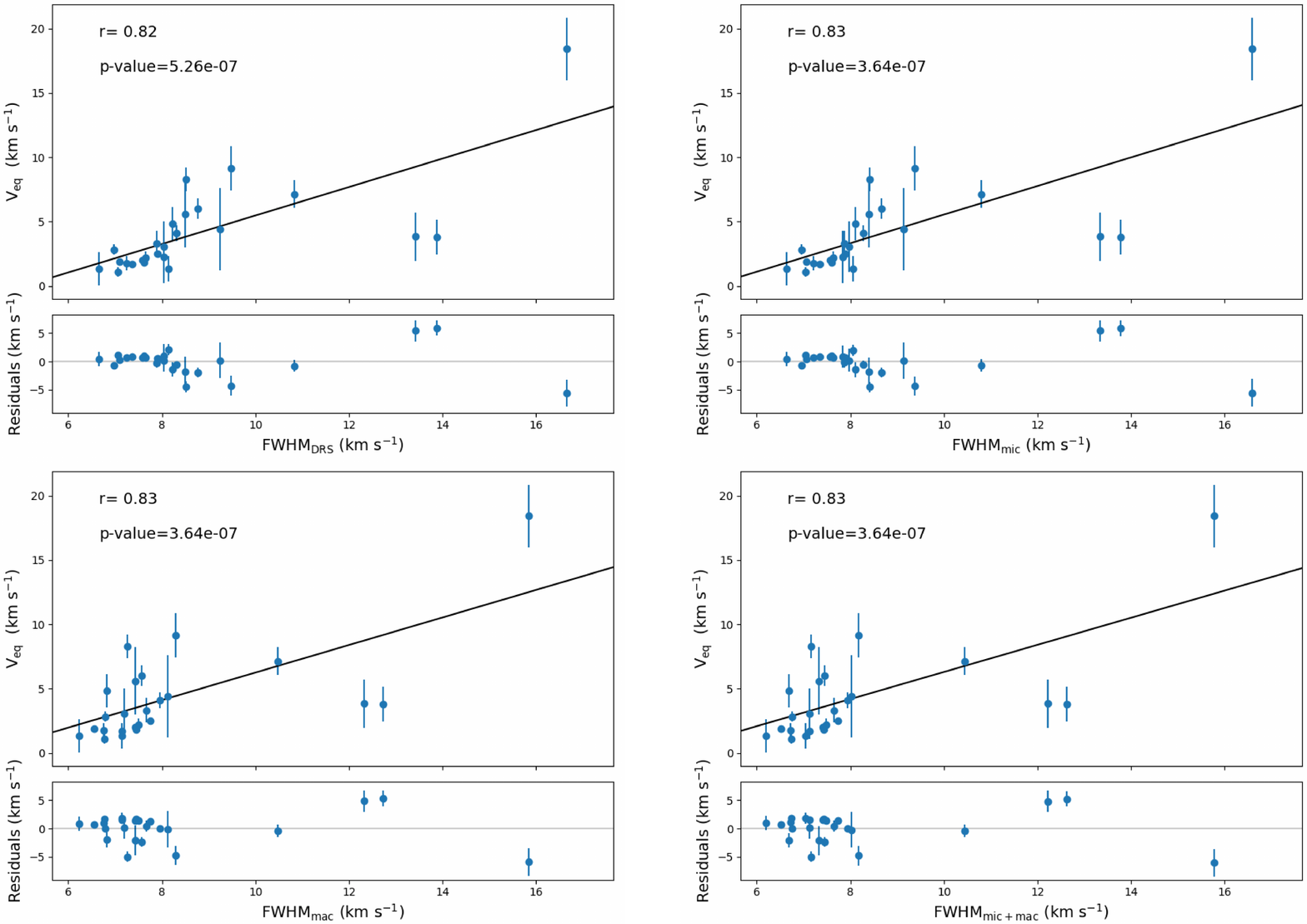}
    \caption{Linear correlations (black solid lines) between the four data sets derived from the FWHM$_\mathrm{DRS}$ computed by the HARPS-N DRS with the G2 mask (\textit{x}-axis) and the stellar equatorial velocity $v_{\mathrm{eq}}$ (\textit{y}-axis) for our set of calibrators. The Spearman's correlation coefficient $r$ and $p$-value are shown in the plots. \textit{Upper left:} Linear correlation between FWHM$_\mathrm{DRS}$ and $v_{\mathrm{eq}}$ and relative residuals. \textit{Upper right:} Linear correlation between FWHM$_\mathrm{mic}$ and $v_{\mathrm{eq}}$ and relative residuals. \textit{Lower left:} Linear correlation between FWHM$_\mathrm{mac}$ and $v_{\mathrm{eq}}$ and relative residuals. \textit{Lower right:} Linear correlation between FWHM$_\mathrm{mic+mac}$ and $v_{\mathrm{eq}}$ and relative residuals.}
    \label{fig:simple}
\end{figure*}

We fitted a linear relation to each one of our four data sets (Fig.~\ref{fig:simple}): \textit{a)} FWHM$_\mathrm{DRS}$ vs. $v_{\mathrm{eq}}$, \textit{b)}  FWHM$_\mathrm{mic}$ vs. $v_{\mathrm{eq}}$, \textit{c)} FWHM$_\mathrm{mac}$ vs. $v_{\mathrm{eq}}$, and \textit{d)}  FWHM$_\mathrm{mic+mac}$ vs. $v_{\mathrm{eq}}$. The three leftmost points (TrES-4, Kepler-25, and HAT-P-8 from lower to higher FWHM respectively) may appear as outliers, but we decided to keep them for several reasons: we have very few calibrators with FWHM > 10 km~s$^{-1}$, we have no solid reason to mistrust the $P_{\mathrm{rot}}$ and $R_{\star}$ values used in our work, and the $v_{\mathrm{eq}}\sin{i_\star}$ computed with the resulting calibrations for hundreds of exoplanet-host stars agree well with the literature values (see Sec.~\ref{sec:vsini}).

We used as final relation the most simple and straightforward one, that links linearly the FWHM$_\mathrm{DRS}$ as it is and the $v_{\mathrm{eq}}\sin{i_\star}$ (Fig.~\ref{fig:simple}, upper left panel), as this is the relation that may be more widely useful, because it does not depend on the knowledge of $v_{\rm micro}$ and $v_{\rm macro}$. The resulting calibrations using the G2 and K5 masks respectively are thus:
\begin{equation}
\label{eq:vsini}
    \begin{split}
        \mathrm{G2\ mask:\ } v_{\mathrm{eq}}\sin{i_\star} = 1.09446 \times \mathrm{FWHM}_\mathrm{DRS} - 5.45380 \\
        \mathrm{K5\ mask:\ } v_{\mathrm{eq}}\sin{i_\star} = 1.26952 \times \mathrm{FWHM}_\mathrm{DRS} - 6.06771 
    \end{split}
\end{equation}

For completeness' sake, we give here also the calibrations obtained for FWHM$_\mathrm{mic}$ (Eq.~\ref{eq:vsini_micro}), FWHM$_\mathrm{mac}$ (Eq.~\ref{eq:vsini_macro}), and FWHM$_\mathrm{mic+mac}$ (Eq.~\ref{eq:vsini_all}):

\begin{equation}
\label{eq:vsini_micro}
    \begin{split}
        \mathrm{G2\ mask:\ } v_{\mathrm{eq}}\sin{i_\star} = 1.09886 \times \mathrm{FWHM}_\mathrm{mic} - 5.42695 \\
        \mathrm{K5\ mask:\ } v_{\mathrm{eq}}\sin{i_\star} = 1.27563 \times \mathrm{FWHM}_\mathrm{mic} - 6.04075 
    \end{split}
\end{equation}

\begin{equation}
\label{eq:vsini_macro}
    \begin{split}
        \mathrm{G2\ mask:\ } v_{\mathrm{eq}}\sin{i_\star} = 1.05962 \times \mathrm{FWHM}_\mathrm{mac} -4.33315 \\
        \mathrm{K5\ mask:\ } v_{\mathrm{eq}}\sin{i_\star} = 1.23413 \times \mathrm{FWHM}_\mathrm{mac} - 4.81863 
    \end{split}
\end{equation}

\begin{equation}
\label{eq:vsini_all}
    \begin{split}
        \mathrm{G2\ mask:\ } v_{\mathrm{eq}}\sin{i_\star} = 1.0438 \times \mathrm{FWHM}_\mathrm{mic+mac} - 4.13~~~~~~~~ \\
        \mathrm{K5\ mask:\ } v_{\mathrm{eq}}\sin{i_\star} = 1.21346 \times \mathrm{FWHM}_\mathrm{mic+mac} - 4.57564
    \end{split}
\end{equation}

To estimate the errors on our $v_{\mathrm{eq}}\sin{i_\star}$ measurements, we applied the error propagation theory. Considering that all our equations are linear fits structured as $v_{\mathrm{eq}}\sin{i_\star} = a\mathrm{FWHM} + b$, we could derive the error on $v_{\mathrm{eq}}\sin{i_\star}$ using the following equation:
\begin{equation}
    \label{eq:simple_error}
    \sigma_{v_{\mathrm{eq}}\sin{i_\star}} = \sqrt{\mathrm{FWHM}^2 \sigma{_a}^2 + \sigma{_b}^2 + 2\mathrm{FWHM}\sigma{_a}\sigma{_b}\rho(a,b)}
\end{equation}
where $\sigma_a$ and $\sigma_b$ are the uncertainties in the fit parameters, while $\rho(a,b)$ is the correlation coefficient:
\begin{equation}
    \rho(a,b) = \frac{- \sum_{i=1}^N FWHM_i}{\sqrt{N\sum_{i=1}^N FWHM_{i}^2}}
\end{equation}
The values of $\sigma{_a}$, $\sigma{_b}$, and $\rho(a,b)$ for all the Eqs.~\ref{eq:vsini},\ref{eq:vsini_micro},\ref{eq:vsini_macro},\ref{eq:vsini_all} are listed in Table~\ref{tab:errors}.
\begin{table*}[]
    \centering
    \caption{Fit parameters $a$ and $b$, uncertainties $\sigma{_a}$ and $\sigma{_b}$, and correlation factor $\rho(a,b)$ for all the relevant equations obtained in this paper.}
    \begin{tabular}{c c c c c c c}
    \hline
    \hline
    Equation & Mask & $a$ & $b$ & $\sigma{_a}$ & $\sigma{_b}$ & $\rho(a,b)$\\
    \hline
    \ref{eq:vsini} & G2 & 1.09446 & -5.45380 & 0.21854 & 2.00007 & -0.96604\\
    \ref{eq:vsini} & K5 & 1.26952 & -6.06771 & 0.19402 & 1.62830 & -0.96250\\
    \ref{eq:vsini_micro} & G2 & 1.09886 & -5.42695 & 0.21989 & 1.9991 & -0.96592 \\
    \ref{eq:vsini_micro} & K5 & 1.27563 & -6.04075 & 0.19522 & 1.62635 & -0.96238 \\
    \ref{eq:vsini_macro} & G2 & 1.05962 & -4.33315 & 0.23369 & 1.96348 & -0.96100 \\
    \ref{eq:vsini_macro} & K5 & 1.23413 & -4.81863 & 0.21586 & 1.64721 & -0.95615 \\
    \ref{eq:vsini_all} & G2 & 1.0438 & -4.13 & 0.23468 & 1.95621 & -0.95998 \\
    \ref{eq:vsini_all} & K5 & 1.21346 & -4.57564 & 0.21850 & 1.65255 & -0.95471 \\
    \ref{eq:vsini_literature} & G2 & 1.1241 & -5.70685 & 0.03629 & 0.32201 & -0.9657\\
    \ref{eq:vsini_lit_k5} & K5 & 1.34470 & -6.69438 & 0.08660 & 0.57299 & -0.99065 \\ 
    \end{tabular}
    \label{tab:errors}
\end{table*}
We have no estimate on the error of $\mathrm{FWHM}_\mathrm{DRS}$, because unfortunately this information is not stored in the header of the FITS files, but we tried to recover it by checking the standard deviation of the FWHM$_\mathrm{DRS}$ values when more than one CCF was available. We found a standard deviation of the order of 4\%, which is much lower than the other contributions to the error budget. Thus we deemed Eq.~\ref{eq:simple_error} sufficient to estimate the errors in $v_{\mathrm{eq}}\sin{i_\star}$ derived from Eq.~\ref{eq:vsini}. Concerning Eqs.~\ref{eq:vsini_micro},~\ref{eq:vsini_macro},~\ref{eq:vsini_all} instead, we need to consider also the error on $v_{\rm micro}$ and $v_{\rm macro}$, that will propagate and give us $\sigma_{\mathrm{FWHM}_\mathrm{mic}}$, $\sigma_{\mathrm{FWHM}_\mathrm{mac}}$, and $\sigma_{\mathrm{FWHM}_\mathrm{mic+mac}}$. The total error then will be:
\begin{equation}
    \sigma_{\mathrm{tot}} = \sqrt{\sigma_{v_{\mathrm{eq}}\sin{i_\star}}^2 + a^2\sigma_{\mathrm{FWHM}}^2}
\end{equation}
As stated before, we used the rms as errors on $v_{\rm micro}$ and $v_{\rm macro}$ (0.18 km~s$^{-1}$ for $v_{\rm micro}$, either 0.5 of 0.73 km~s$^{-1}$ for $v_{\rm micro}$ depending of the star's temperature -- the former for $T_{\mathrm{eff}} < 5700$ K, the latter for $T_{\mathrm{eff}} > 5700$ K). These values are larger than what we would obtain propagating the errors on the stellar parameters.

We compared the results obtained with the different calibration on our calibrators set (see Table~\ref{tab:comparison}), and the $v_{\mathrm{eq}}\sin{i_\star}$ agree to the order of 0.2-0.3 km~s$^{-1}$ with the exception of WASP-14, where Eqs.~\ref{eq:vsini} and \ref{eq:vsini_micro} give very different results from Eqs.~\ref{eq:vsini_macro} and \ref{eq:vsini_all}: WASP-14 is the hottest star in our calibrators' set, with the largest $v_{\rm micro}$ and $v_{\rm macro}$ values, and the problems may arise from an over-estimating of these values due to the stellar $T_{\mathrm{eff}}$ being at the edge of the applicability range of the relationships used to compute them.
%

\begin{table*}
    \caption{Comparison between $v_{\mathrm{eq}}\sin{i_\star}$ obtained with the different Eqs.~\ref{eq:vsini}, \ref{eq:vsini_micro}, \ref{eq:vsini_micro}, \ref{eq:vsini_all} for our calibrators, along with the standard deviation of the results. The spectral types are taken from SIMBAD.}
    \label{tab:comparison}
    \centering
    \begin{tabular}{c c c c c c c c}
    \hline
    \hline 
    Name & $v_{\mathrm{eq}}\sin{i_\star}$ & $v_{\mathrm{eq}}\sin{i_\star}_\mathrm{mic}$ & $v_{\mathrm{eq}}\sin{i_\star}_\mathrm{mac}$ & $v_{\mathrm{eq}}\sin{i_\star}_\mathrm{mic+mac}$ & Std. dev. & Mask used & Sp. type\\
    & [km~s$^{-1}$] & [km~s$^{-1}$] & [km~s$^{-1}$] & [km~s$^{-1}$] & [km~s$^{-1}$] &  for the CCF & \\
    \hline
    HAT-P-1 & 3.46$\pm0.54$ & 3.43$\pm0.54$ & 3.23$\pm0.72$ & 3.22$\pm0.73$ & 0.11 & G2 & G0V \\
     & 3.81$\pm0.45$ & 3.76$\pm0.45$ & 3.47$\pm0.72$ & 3.45$\pm0.73$ & 0.16 & K5 & G0V \\
    HAT-P-3 & 2.28$\pm0.65$ & 2.31$\pm0.64$ & 2.84$\pm0.64$ & 2.91$\pm0.64$ & 0.29 & G2 & K1V \\
     & 1.88$\pm0.57$ & 1.91$\pm0.59$ & 2.49$\pm0.6$ & 2.58$\pm0.60$ & 0.32 & K5 & K1V\\
    HAT-P-8 & 12.77$\pm1.78$ & 12.80$\pm1.79$ & 12.45$\pm1.91$ & 12.33$\pm1.92$ & 0.20 & G2 & F8V\\
     & 15.55$\pm1.65$ & 15.6$\pm1.66$ & 15.22$\pm1.85$ & 15.05$\pm1.87$ & 0.23 & K5 & F8V\\
    HAT-P-13 & 3.34$\pm0.55$ & 3.34$\pm0.55$ & 3.30$\pm0.64$ & 3.31$\pm0.64$ & 0.01 & G2 & G4\\
    HAT-P-16 & 3.54$\pm0.53$ & 3.48$\pm0.54$ & 2.90$\pm0.81$ & 2.85$\pm0.82$ & 0.32 & G2 & F8\\
     & 3.91$\pm0.44$ & 3.83$\pm0.45$ & 3.06$\pm0.83$ & 2.99$\pm0.84$ & 0.42 & K5 & F8\\
    HAT-P-17 & 1.84$\pm0.70$ & 1.86$\pm0.70$ & 2.27$\pm0.72$ & 2.34$\pm0.72$ & 0.23 & G2 & G0\\
     & 1.75$\pm0.58$ & 1.77$\pm0.58$ & 2.20$\pm0.65$ & 2.29$\pm0.65$ & 0.25 & K5 & G0\\
    HAT-P-20 & 3.20$\pm0.56$ & 3.24$\pm0.55$ & 3.88$\pm0.56$ & 3.95$\pm0.56$ & 0.35 & G2 & K3V\\
     & 2.92$\pm0.48$ & 2.97$\pm0.48$ & 3.71$\pm0.51$ & 3.79$\pm0.51$ & 0.40 & K5 & K3V\\
    HAT-P-22 & 2.31$\pm0.64$ & 2.33$\pm0.64$ & 2.62$\pm0.68$ & 2.68$\pm0.68$ & 0.17 & G2 & G5\\
     & 2.26$\pm0.53$ & 2.28$\pm0.53$ & 2.55$\pm0.63$ & 2.62$\pm0.63$ & 0.16 & K5 & G5\\
    HD17156 & 4.14$\pm0.52$ & 4.10$\pm0.52$ & 3.68$\pm0.72$ & 3.64$\pm0.72$ & 0.23 & G2 & F9V\\
     & 4.61$\pm0.45$ & 4.56$\pm0.45$ & 4.00$\pm0.74$ & 3.95$\pm0.75$ & 0.31 & K5 & F9V\\
    HD63433 & 6.40$\pm0.67$ & 6.43$\pm0.68$ & 6.77$\pm0.79$ & 6.77$\pm0.80$ & 0.18 & G2 & G5V\\
     & 7.23$\pm0.64$ & 7.28$\pm0.65$ & 7.66$\pm0.79$ & 7.65$\pm0.81$ & 0.20 & K5 & G5V\\
    HD189733 & 3.18$\pm0.56$ & 3.22$\pm0.56$ & 3.79$\pm0.57$ & 3.85$\pm0.57$ & 0.31 & G2 & K2V\\
     & 3.11$\pm0.47$ & 3.15$\pm0.47$ & 3.80$\pm0.51$ & 3.87$\pm0.52$ & 0.35 & K5 & K2V\\
    HD209458 & 3.84$\pm0.52$ & 3.80$\pm0.53$ & 3.55$\pm0.71$ & 3.52$\pm0.71$ & 0.14 & G2 & F9V\\
     & 4.30$\pm0.44$ & 4.25$\pm0.44$ & 3.89$\pm0.71$ & 3.85$\pm0.72$ & 0.20 & K5 & F9V\\
    K2-29 & 3.64$\pm0.53$ & 3.67$\pm0.53$ & 4.10$\pm0.57$ & 4.15$\pm0.57$ & 0.24 & G2 & K2V\\
     & 3.79$\pm0.45$ & 3.82$\pm0.45$ & 4.30$\pm0.52$ & 4.35$\pm0.53$ & 0.26 & K5 & K2V\\
    K2-34 & 4.91$\pm0.53$ & 4.88$\pm0.53$ & 4.45$\pm0.69$ & 4.41$\pm0.70$ & 0.23 & G2 & G2V\\
     & 5.59$\pm0.49$ & 5.55$\pm0.49$ & 5.01$\pm0.72$ & 4.95$\pm0.73$ & 0.30 & K5 & G2V\\
    Kepler-25 & 9.73$\pm1.21$ & 9.71$\pm1.21$ & 9.15$\pm1.26$ & 9.04$\pm1.26$ & 0.32 & G2 & --\\
    Qatar-1 & 2.61$\pm0.61$ & 2.65$\pm0.61$ & 3.23$\pm0.60$ & 3.30$\pm0.60$ & 0.32 & G2 & --\\
     & 2.28$\pm0.53$ & 2.32$\pm0.53$ & 2.98$\pm0.56$ & 3.07$\pm0.56$ & 0.36 & K5 & --\\
    Qatar-2 & 2.85$\pm0.58$ & 2.90$\pm0.58$ & 3.54$\pm0.57$ & 3.61$\pm0.58$ & 0.35 & G2 & K5V\\
     & 2.45$\pm0.52$ & 2.50$\pm0.51$ & 3.24$\pm0.53$ & 3.32$\pm0.54$ & 0.40 & K5 & K5V\\
    TrES-4 & 9.23$\pm1.12$ & 9.22$\pm1.13$ & 8.72$\pm1.18$ & 8.63$\pm1.18$ & 0.28 & G2 & --\\
     & 10.55$\pm1.05$ & 10.54$\pm1.05$ & 9.94$\pm1.17$ & 9.82$\pm1.17$ & 0.34 & K5 & --\\
    WASP-11 & 2.18$\pm0.66$ & 2.22$\pm0.65$ & 2.85$\pm0.63$ & 2.93$\pm0.63$ & 0.35 & G2 & K3V\\
     & 1.84$\pm0.57$ & 1.87$\pm0.57$ & 2.59$\pm0.58$ & 2.69$\pm0.59$ & 0.39 & K5 & K3V\\
    WASP-13 & 3.86$\pm0.52$ & 3.82$\pm0.52$ & 3.36$\pm0.74$ & 3.34$\pm0.75$ & 0.25 & G2 & G1V\\
     & 4.30$\pm0.44$ & 4.25$\pm0.44$ & 3.64$\pm0.76$ & 3.60$\pm0.77$ & 0.33 & K5 & G1V\\
    WASP-14 & 3.35$\pm0.55$ & 3.18$\pm0.56$ & 0.61$\pm1.45$ & 0.35$\pm1.55$ & 1.40 & G2 & F5V\\
     & 3.75$\pm0.45$ & 3.54$\pm0.46$ & 0.25$\pm1.66$ & 0.00$\pm1.83$ & 1.79 & K5 & F5V\\
    WASP-32 & 4.66$\pm0.52$ & 4.62$\pm0.52$ & 4.28$\pm0.68$ & 4.24$\pm0.69$ & 0.19 & G2 & --\\
     & 5.28$\pm0.47$ & 5.23$\pm0.47$ & 4.79$\pm0.71$ & 4.73$\pm0.71$ & 0.25 & K5 & --\\
    WASP-43 & 2.91$\pm0.58$ & 2.96$\pm0.58$ & 3.62$\pm0.57$ & 3.68$\pm0.57$ & 0.36 & G2 & K7V\\
     & 2.68$\pm0.50$ & 2.73$\pm0.50$ & 3.50$\pm0.51$ & 3.58$\pm0.52$ & 0.42 & K5 & K7V\\
    WASP-69 & 2.87$\pm0.58$ & 2.92$\pm0.58$ & 3.55$\pm0.57$ & 3.62$\pm0.58$ & 0.35 & G2 & --\\
     & 2.51$\pm0.51$ & 2.56$\pm0.51$ & 3.29$\pm0.53$ & 3.38$\pm0.53$ & 0.40 & K5 & --\\
    WASP-84 & 2.97$\pm0.48$ & 3.00$\pm0.48$ & 3.35$\pm0.56$ & 3.41$\pm0.57$ & 0.20 & K5 & --\\
    XO-2N & 2.47$\pm0.62$ & 2.49$\pm0.62$ & 2.83$\pm0.66$ & 2.89$\pm0.66$ & 0.19 & G2 & G9V\\
     & 2.27$\pm0.53$ & 2.28$\pm0.53$ & 2.63$\pm0.61$ & 2.69$\pm0.62$ & 0.19 & K5 & G9V\\
    \hline
    \end{tabular}
\end{table*}

%


\section{Projected rotational velocity of exoplanet-host stars}\label{sec:vsini}

We decided to apply our relation to all the HARPS-N observed exoplanet-host stars found in the TNG archive. First, we queried again the NASA exoplanet archive to obtain a complete list of all known exoplanet-host stars with declination $> -25\deg$, without any other constraints. We obtained a preliminary list of 3750 exoplanets (2753 host stars).

We queried the TNG archive\footnote{\url{http://archives.ia2.inaf.it/tng/}} with a self-written \texttt{python} code using the  \texttt{pyvo} module\footnote{\url{https://pyvo.readthedocs.io/en/latest/index.html}} in an asynchronous TAP query, retrieving up to 10 public CCF FITS files for each target. We found data for 313 stars, but some of them are useless for different reasons, e.g. fast rotating stars, too low signal-to-noise ratio (SNR), M-type stars reduced with the M2 mask.

We point out here that the CCFs of M-type stars may be used if they are computed with the K5 mask: this results in a noisier, but more physically significant CCF. We were also able to recover the M-type stars reduced with the M2 mask that were observed within the GAPS program: in this case, we could reduce again the spectra with the K5 mask using the YABI platform \citep{Hunter2012} hosted at the IA2 Data Center\footnote{\url{https://www.ia2.inaf.it}}.

In the end, we had to discard some non-GAPS stars having only M2-mask public CCFs, and others stars whose CCFs had too low SNR, or the wrong input radial velocity. We estimated the $v_{\mathrm{eq}}\sin{i_\star}$ for all the 273 remaining targets with FWHM$_\mathrm{DRS}$ < 20 km~s$^{-1}$. Our $v_{\mathrm{eq}}\sin{i_\star}$ values are reported in Table~\ref{tab:all_vsini}, the errors are computed using Eq.~\ref{eq:simple_error}.
\longtab{
\begin{longtable}{c c c c c c c}
    \caption{Computed $v_{\mathrm{eq}}\sin{i_\star}$ of exoplanet-host stars. When more than one spectrum is found in the archive, the FWHM$_\mathrm{DRS}$ is obtained as the mean of a maximum of 10 values. The spectral types are taken from SIMBAD.}
    \label{tab:all_vsini}
    \centering
    & & & & & This work & Literature \\
    \hline
    \hline \\[-6pt] 
    Name & Sp. Type & DRS mask & FWHM$_\mathrm{DRS}$ & $v_{\mathrm{eq}}\sin{i_\star}$ & Lit. $v_{\mathrm{eq}}\sin{i_\star}$ & Reference \\ 
    & & & (km~s$^{-1}$) & (km~s$^{-1}$) & (km~s$^{-1}$) &  \\ 
    \hline \\[-6pt] 
    \endfirsthead
    \caption{continued.}\\
    & & & & This work & Literature & \\
    \hline
    \hline
    Name & Sp. Type & DRS mask & FWHM$_\mathrm{DRS}$ & $v_{\mathrm{eq}}\sin{i_\star}$ & Lit. $v_{\mathrm{eq}}\sin{i_\star}$ & Reference \\ 
    & & & (km~s$^{-1}$) & (km~s$^{-1}$) & (km~s$^{-1}$) &  \\
    \hline
    \endhead
    \hline
    \endfoot
        2MASS J22362452 & -- & K5 & 7.44 & 3.38$\pm$0.46 & -- & -- \\
        +4751425 & & & & & \\
        24 Sex & K0IV & G2 & 7.26 & 2.49$\pm$0.62 &  2.77$\pm$0.5  &  \cite{Johnson2011b} \\ 
        51 Peg & G2IV & G2 & 7.31 & 2.54$\pm$0.62 &  2.2$\pm$1.0  &  \cite{Mayor1995} \\ 
        55 Cnc & K0IV-V & G2 & 7.08 & 2.30$\pm$0.64 &  2.0$\pm$0.0  &  \cite{Butler1997} \\ 
        BD+03 2562 & K2 & K5 & 8.05 & 4.15$\pm$0.44 &  2.7$\pm$0.3  &  \cite{Villaver2017} \\ 
        BD+14 4559 & K5 & K5 & 6.24 & 1.85$\pm$0.57 &  2.5$\pm$1.0  &  \cite{Niedzielski2009} \\ 
        BD+15 2375 & -- & K5 & 7.67 & 3.66$\pm$0.45 &  2.5$\pm$0.5  &  \cite{Niedzielski2016} \\ 
        BD+20 2457 & K2 & K5 & 8.09 & 4.21$\pm$0.44 &  2.5$\pm$0.0  &  \cite{Niedzielski2009} \\ 
        BD+48 740 & K3III & K5 & 8.27 & 4.43$\pm$0.44 &  0.7$\pm$2.2  &  \cite{Adamow2018} \\ 
        BD-11 4672 & K7V & K5 & 6.15 & 1.74$\pm$0.58 &  1.0$\pm$0.5  &  \cite{Barbato2020} \\ 
        CI Tau & K4IVe & K5 & 15.93 & 14.16$\pm$1.59 &  --  &  -- \\ 
        CoRoT-7 & K0V & G2 & 7.10 & 2.31$\pm$0.64 &  <3.5  &  \cite{Leger2009} \\ 
        CoRoT-24 & -- & K5 & 6.34 & 1.99$\pm$0.56 & 2.0$\pm$1.5  &  \cite{Alonso2014} \\ 
        EPIC 211945201 & -- & G2 & 7.66 & 2.93$\pm$0.58 &  4.0$\pm$1.0  &  \cite{Chakraborty2018} \\ 
        EPIC 220674823 & -- & G2 & 6.83 & 2.02$\pm$0.68 &  2.62$\pm$0.18  &  \cite{Adams2017} \\ 
        EPIC 249893012 & -- & G2 & 7.12 & 2.34$\pm$0.64 &  2.1$\pm$0.5  &  \cite{Hidalgo2020} \\ 
        GJ 15 A & M2V & K5 & 5.37 & 0.76$\pm$0.68 &  1.09$\pm$0.79  &  \cite{Pinamonti2018} \\ 
        GJ 328 & M0V & K5 & 6.00 & 1.55$\pm$0.60 & --  &  -- \\ 
        GJ 338 B & M0V & K5 & 6.15 & 1.74$\pm$0.58 &  2.3$\pm$1.5  &  \cite{Gonzalez2020} \\ 
        GJ 357 & M2.5V & G2 & 5.71 & 0.80$\pm$0.86 &  <2.0  &  \cite{Reiners2018} \\ 
        GJ 436 & M3V & K5 & 5.11 & 0.41$\pm$0.73 & 0.33$\pm$0.09  &  \cite{Bourrier2018} \\ 
        GJ 504 & G0V  & G2 & 11.19 & 6.79$\pm$0.73 & 6.0$\pm$1.0  &  \cite{DOrazi2017} \\ 
        GJ 625 & M1.5V & K5 & 5.50 & 0.92$\pm$0.18 & 0.2$\pm$0.0  &  \cite{Suarez2017} \\ 
        GJ 685 & M1V & K5 & 5.89 & 1.41$\pm$0.61 & 1.33$\pm$0.42  &  \cite{Pinamonti2019} \\ 
        GJ 3470 & M2.0V & K5 & 5.50 & 0.92$\pm$0.67 &  --  &  -- \\ 
        GJ 3942 & M0.5V  & K5 & 6.08 & 1.65$\pm$0.59 &  1.67$\pm$0.03  &  \cite{Perger2017} \\ 
        GJ 3998 & M1V & K5 & 5.55 & 0.98$\pm$0.66 &  0.93$\pm$0.55 &  \cite{Affer2016} \\ 
        GJ 9827 & K6V & K5 & 6.12 & 1.70$\pm$0.58 &  2.0$\pm$0.0  &  \cite{Rice2019} \\ 
        Gl 49 & M1.5V & K5 & 5.83 & 1.34$\pm$0.62 &  2.0$\pm$0.0  & \cite{Perger2019} \\ 
        Gl 686 & M1.5V & K5 & 5.38 & 0.76$\pm$0.68 &  1.01$\pm$0.80  &  \cite{Affer2019} \\ 
        HAT-P-1 & G0V & G2 & 8.14 & 3.46$\pm$0.54 &  3.75$\pm$0.58  &  \cite{Johnson2008b} \\
        HAT-P-3 & K1V & K5 & 6.26 & 1.88$\pm$0.57 &  1.4$\pm$0.5  &  \cite{Biazzo2022} \\ 
        HAT-P-4 & F & G2 & 9.64 & 5.10$\pm$0.55 &  5.6$\pm$0.5  &  \cite{Biazzo2022} \\ 
        HAT-P-6 & F8V & G2 & 12.37 & 8.09$\pm$0.93 &  7.8$\pm$0.6  &  \cite{Albrecht2012} \\ 
        HAT-P-7 & F6V & G2 & 9.22 & 4.63$\pm$0.52 & 3.8$\pm$0.5  &  \cite{Bonomo2017} \\ 
        HAT-P-8 & F8V & G2 & 17.31 & 12.77$\pm$1.78 &  11.5$\pm$0.5  &  \cite{Bonomo2017} \\ 
        HAT-P-12 & K5 & K5 & 6.01 & 1.56$\pm$0.60 &  0.5$\pm$0.5  &  \cite{Biazzo2022} \\ 
        HAT-P-13 & G4 & G2 & 8.04 & 3.34$\pm$0.55 &  2.90$\pm$1.0  &  \cite{Bonomo2017} \\
        HAT-P-14 & F5V & G2 & 12.53 & 8.26$\pm$0.96 &  8.1$\pm$0.3  &  \cite{Biazzo2022} \\
        HAT-P-15 & G5 & G2 & 7.19 & 2.42$\pm$0.63 &  2.0$\pm$0.3  &  \cite{Biazzo2022} \\ 
        HAT-P-16 & F8 & G2 & 8.22 & 3.54$\pm$0.53 & 3.5$\pm$0.5  &  \cite{Buchhave2010} \\ 
        HAT-P-17 & G0 & G2 & 6.66 & 1.84$\pm$0.70 &  1.3$\pm$0.3  & \cite{Biazzo2022} \\ 
        HAT-P-18 & K2V & K5 & 6.26 & 1.88$\pm$0.57 &  1.3$\pm$0.4  &  \cite{Biazzo2022} \\ 
        HAT-P-20 & K3V & K5 & 7.08 & 2.92$\pm$0.48 &  2.6$\pm$0.3  &  \cite{Biazzo2022} \\ 
        HAT-P-21 & G3 & G2 & 8.20 & 3.52$\pm$0.54 & 3.9$\pm$0.5  &  \cite{Biazzo2022} \\ 
        HAT-P-22 & G5 & G2 & 7.12 & 2.31$\pm$0.64 & 1.8$\pm$0.5  &  \cite{Biazzo2022} \\ 
        HAT-P-24 & F8 & G2 & 15.0 & 10.96$\pm$1.44 &  10.0$\pm$0.5  &  \cite{Bonomo2017} \\
        HAT-P-26 & K1 & K5 & 6.01 & 1.56$\pm$0.60 &  0.8$\pm$0.4  &  \cite{Biazzo2022} \\ 
        HAT-P-29 & -- & G2 & 9.11 & 4.52$\pm$0.52 &  4.5$\pm$0.7  &  \cite{Biazzo2022} \\ 
        HAT-P-30 & G0 & G2 & 8.38 & 3.72$\pm$0.53 & 3.0$\pm$0.5  &  \cite{Biazzo2022} \\ 
        HAT-P-31 & -- & G2 & 7.55 & 2.81$\pm$0.59 &  0.5$\pm$0.6 &  \cite{Bonomo2017} \\ 
        HAT-P-36 & -- & G2 & 7.90 & 3.19$\pm$0.56 &  3.6$\pm$0.3  &  \cite{Biazzo2022} \\ 
        HAT-P-44 & -- & K5 & 6.32 & 1.96$\pm$0.56 &  0.2$\pm$0.5  &  \cite{Hartman2014} \\ 
        HAT-P-46 & -- & K5 & 8.93 & 5.27$\pm$0.47 &  4.9$\pm$0.5  &  \cite{Hartman2014} \\ 
        HAT-P-49 & A8 & G2 & 17.73 & 13.95$\pm$2.01 &  16.0$\pm$0.5  &  \cite{Bonomo2017} \\ 
        HAT-P-50 & -- & G2 & 11.90 & 7.57$\pm$0.84 &  8.9$\pm$0.5  &  \cite{Hartman2015} \\ 
        HAT-P-54 & -- & K5 & 7.05 & 2.88$\pm$0.48 &  2.35$\pm$0.5  &  \cite{Bakos2015} \\ 
        HD 3167 & K0V & K5 & 6.17 & 1.77$\pm$0.58 &  1.7$\pm$1.1  &  \cite{Christiansen2017} \\ 
        HD 3651 & K0.5V & K5 & 6.20 & 1.80$\pm$0.57 &  1.1$\pm$0.0  &  \cite{Wittenmyer2009} \\ 
        HD 5583 & K0 & K5 & 8.43 & 4.63$\pm$0.45 &  2.2$\pm$2.3  &  \cite{Niedzielski2016} \\ 
        HD 11506 & G0V & G2 & 9.58 & 5.03$\pm$0.54 &  5.0$\pm$0.5  &  \cite{Fischer2007} \\ 
        HD 13931 & G0 & G2 & 7.18 & 2.41$\pm$0.63 &  --  &  -- \\ 
        HD 17156 & F9V & G2 & 8.78 & 4.14$\pm$0.52 &  2.8$\pm$0.5  &  \cite{Bonomo2017} \\ 
        HD 23596 & F8 & G2 & 8.27 & 3.60$\pm$0.53 &  4.2$\pm$0.0  &  \cite{Wittenmyer2009} \\ 
        HD 26965 & K0V & K5 & 5.92 & 1.44$\pm$0.61 &  --  &  -- \\ 
        HD 32963 & G5IV & G2 & 6.88 & 2.07$\pm$0.67 &  --  &  -- \\ 
        HD 38801 & G8IV & K5 & 6.84 & 2.62$\pm$0.50 &  --  &  -- \\ 
        HD 46375 & G9V & K5 & 6.22 & 1.82$\pm$0.57 &  0.86$\pm$0.0  &  \cite{Butler2006} \\ 
        HD 50554 & F8V & G2 & 7.99 & 3.29$\pm$0.55 &  3.88$\pm$0.0  &  \cite{Fischer2002} \\ 
        HD 63433 & G5V & G2 & 10.73 & 6.40$\pm$0.67 &  7.28$\pm$0.29  &  \citep{Mann2020} \\
        HD 68988 & G1V & G2 & 7.89 & 3.18$\pm$0.56 &  2.84$\pm$0.0  &  \cite{Butler2006} \\ 
        HD 72659 & G2V & G2 & 7.22 & 2.45$\pm$0.63 &  2.2$\pm$0.5  &  \cite{Valenti2005} \\ 
        HD 73534 & G5 & G2 & 7.10 & 2.31$\pm$0.64 &  1.72$\pm$0.29  &  \cite{Jofre2015} \\ 
        HD 75898 & G0 & G2 & 8.58 & 3.94$\pm$0.52 &  4.54$\pm$0.5  &  \cite{Robinson2007} \\ 
        HD 80606 & G8V & G2 & 7.13 & 2.36$\pm$0.64 &  1.8$\pm$0.5  &  \cite{Bonomo2017} \\ 
        HD 87883 & K0V & K5 & 6.11 & 1.69$\pm$0.58 &  2.17$\pm$0.5  &  \cite{Fischer2009} \\ 
        HD 88133 & G8V & G2 & 7.30 & 2.54$\pm$0.62 &  2.2$\pm$0.3  &  \cite{Fischer2005} \\ 
        HD 89307 & G0V & G2 & 7.38 & 2.62$\pm$0.61 &  3.21$\pm$0.5  &  \cite{Fischer2009} \\ 
        HD 89345 & G5 & G2 & 7.60 & 2.86$\pm$0.58 &  2.6$\pm$0.5  &  \cite{VanEylen2018} \\ 
        HD 96063 & G6V & K5 & 6.61 & 2.32$\pm$0.53 & 0.87$\pm$0.5  &  \cite{Johnson2011a} \\ 
        HD 99109 & G8/K0IV & K5 & 6.40 & 2.06$\pm$0.55 &  1.86$\pm$0.0  &  \cite{Butler2006} \\ 
        HD 102272 & K2 & K5 & 8.50 & 4.72$\pm$0.45 &  3.0$\pm$1.0  &  \cite{Niedzielski2009b} \\ 
        HD 106252 & G0 & G2 & 6.99 & 2.20$\pm$0.66 &  1.9$\pm$0.0  &  \cite{Wittenmyer2009} \\ 
        HD 106515 A & G5V & G2 & 6.69 & 1.86$\pm$0.70 &  <1.0  &  \cite{Marmier2013} \\ 
        HD 107148 & G5V & K5 & 6.68 & 2.42$\pm$0.52 &  0.73$\pm$0.50  &  \cite{Butler2006} \\ 
        HD 108874 & G9V & G2 & 7.02 & 2.22$\pm$0.65 &  2.2$\pm$0.0  &  \cite{Wittenmyer2009} \\ 
        HD 118203 & G0V & K5 & 8.93 & 5.26$\pm$0.47 &  4.7$\pm$0.0  &  \cite{daSilva2006} \\ 
        HD 132406 & G0V & G2 & 7.16 & 2.38$\pm$0.63 &  --  &  -- \\ 
        HD 136418 & G5 & K5 & 6.50 & 2.18$\pm$0.54 &  --  &  -- \\ 
        HD 142245 & K0 & K5 & 6.74 & 2.49$\pm$0.51 &  2.66$\pm$0.5  &  \cite{Johnson2011a} \\ 
        HD 143105 & F5 & G2 & 14.35 & 10.26$\pm$1.31 &  9.1$\pm$1.0  &  \cite{Hebrard2016} \\ 
        HD 149026 & G0IV & G2 & 10.3 & 5.82$\pm$0.61 &  6.0$\pm$0.5  &  \cite{Bonomo2017} \\ 
        HD 149143 & G3V & G2 & 8.36 & 3.70$\pm$0.53 &  3.9$\pm$0.0  &  \cite{daSilva2006} \\ 
        HD 155358 & G0 & G2 & 6.74 & 1.93$\pm$0.69 &  --  &  -- \\ 
        HD 164922 & G9V & K5 & 6.32 & 1.96$\pm$0.56 &  2.0$\pm$0.0  &  \cite{Benatti2020} \\ 
        HD 170469 & G5 & G2 & 7.63 & 2.90$\pm$0.58 &  1.7$\pm$0.5  &  \cite{Fischer2007} \\ 
        HD 176986 & K2.5V & K5 & 6.23 & 1.84$\pm$0.57 &  1.1$\pm$0.2  &  \cite{Suarez2018} \\ 
        HD 183263 & G6IV & G2 & 7.70 & 2.98$\pm$0.57 &  --  &  -- \\ 
        HD 187123 & G2V & G2 & 7.10 & 2.31$\pm$0.64 &  --  &  -- \\ 
        HD 188015 & G5IV & G2 & 7.25 & 2.49$\pm$0.62 &  --  &  -- \\ 
        HD 189733 & K2V & K5 & 7.22 & 3.11$\pm$0.47 &  3.5$\pm$1.0  &  \cite{Bonomo2017} \\ 
        HD 190007 & K5V & K5 & 6.30 & 1.93$\pm$0.56 &  2.55$\pm$0.0  &  \cite{Burt2021} \\ 
        HD 191939 & G0 & G2 & 6.54 & 1.71$\pm$0.72 &  0.6$\pm$0.5  &  \cite{Badenas2020} \\ 
        HD 204941 & K2V & K5 & 6.00 & 1.55$\pm$0.60 &  --  &  -- \\ 
        HD 208897 & K0 & G2 & 7.39 & 2.64$\pm$0.61 &  3.9$\pm$0.42  &  \cite{Yilmaz2017} \\ 
        HD 209458 & F9V & G2 & 8.49 & 3.84$\pm$0.52 &  4.4$\pm$0.2  &  \cite{Albrecht2012} \\ 
        HD 210277 & G8V & G2 & 6.92 & 2.12$\pm$0.67 &  --  &  -- \\ 
        HD 216536 & K0 & K5 & 7.96 & 4.04$\pm$0.44 &  2.6$\pm$0.5  &  \cite{Niedzielski2015b} \\ 
        HD 217786 & F8V & G2 & 7.20 & 2.43$\pm$0.63 &  1.4$\pm$0.0  &  \cite{Moutou2011} \\ 
        HD 218566 & K3V & K5 & 6.23 & 1.84$\pm$0.57 &  --  &  -- \\ 
        HD 219134 & K3V & K5 & 6.07 & 1.64$\pm$0.59 &  0.4$\pm$0.5  &  \cite{Motalebi2015} \\ 
        HD 219828 & G0IV & G2 & 7.66 & 2.93$\pm$0.58 &  2.9$\pm$0.0  &  \cite{Santos2016} \\ 
        HD 220197 & G5 & G2 & 6.40 & 1.55$\pm$0.74 &  1.5$\pm$0.5  &  \cite{Barbato2019} \\ 
        HD 220773 & G0 & G2 & 8.23 & 3.55$\pm$0.53 &  --  &  -- \\ 
        HD 233832 & K2 & K5 & 5.66 & 1.12$\pm$0.64 &  0.8$\pm$0.5  &  \cite{Barbato2019} \\ 
        HD 238914 & K7 & K5 & 7.75 & 3.77$\pm$0.45 &  2.5$\pm$0.5  &  \cite{Adamow2018} \\ 
        HD 285507 & K4 & K5 & 7.19 & 3.07$\pm$0.47 &  3.2$\pm$0.5  &  \cite{Quinn2014} \\ 
        HD 290327 & G0 & G2 & 6.66 & 1.84$\pm$0.70 &  1.44$\pm$1.0  &  \cite{Naef2010} \\ 
        HD 332231 & F8 & G2 & 9.61 & 5.06$\pm$0.54 &  --  &  -- \\ 
        HIP 41378 & F8 & G2 & 9.80 & 5.27$\pm$0.56 &  7.13$\pm$0.5  &  \cite{Vanderburg2016} \\ 
        K2-3 & M1V & K5 & 5.74 & 1.21$\pm$0.63 &  --  &  -- \\ 
        K2-10 & -- & G2 & 6.81 & 2.00$\pm$0.68 &  3.0$\pm$1.0  &  \cite{VanEylen2016} \\ 
        K2-12 & -- & G2 & 7.07 & 2.29$\pm$0.65 &  --  &  -- \\ 
        K2-19 & -- & G2 & 7.09 & 2.31$\pm$0.64 &  2.0$\pm$0.0  &  \cite{Sinukoff2016} \\ 
        K2-27 & -- & K5 & 6.21 & 1.82$\pm$0.57 &  2.3$\pm$0.0  &  \cite{Petigura2017} \\ 
        K2-29 & K2V & K5 & 7.75 & 3.79$\pm$0.45 &  3.7$\pm$0.5  &  \cite{Santerne2016} \\ 
        K2-30 & -- & G2 & 6.88 & 2.08$\pm$0.67 &  1.4$\pm$0.3  &  \cite{Johnson2016} \\ 
        K2-34 & G2V & G2 & 9.47 & 4.91$\pm$0.53 &  6.31$\pm$0.2  &  \cite{Brahm2016} \\ 
        K2-36 & -- & K5 & 6.60 & 2.31$\pm$0.53 &  2.0$\pm$0.0  &  \cite{Damasso2019} \\ 
        K2-39 & -- & K5 & 6.86 & 2.64$\pm$0.50 &  2.01$\pm$0.5  &  \cite{VanEylen2016AJ} \\ 
        K2-44 & -- & G2 & 7.79 & 3.07$\pm$0.57 &  --  &  -- \\ 
        K2-46 & -- & G2 & 8.92 & 4.31$\pm$0.52 &  --  &  -- \\ 
        K2-58 & -- & K5 & 6.55 & 2.25$\pm$0.53 &  --  &  -- \\ 
        K2-60 & -- & G2 & 7.06 & 2.27$\pm$0.65 &  2.2$\pm$0.5  &  \cite{Eigmuller2017} \\ 
        K2-79 & -- & G2 & 7.23 & 2.46$\pm$0.63 &  --  &  -- \\ 
        K2-87 & K1V & K5 & 6.64 & 2.36$\pm$0.52 &  --  &  -- \\ 
        K2-98 & -- & G2 & 10.33 & 5.85$\pm$0.61 &  6.1$\pm$0.5  &  \cite{Barragan2016} \\ 
        K2-99 & -- & G2 & 14.58 & 10.50$\pm$1.36 &  9.3$\pm$0.5  &  \cite{Smith2017} \\ 
        K2-101 & K3V & K5 & 7.62 & 3.61$\pm$0.45 &  3.9$\pm$0.9  &  \cite{Mann2017} \\ 
        K2-105 & -- & G2 & 7.08 & 2.29$\pm$0.64 &  1.76$\pm$0.86  &  \cite{Narita2017} \\ 
        K2-108 & -- & G2 & 7.28 & 2.51$\pm$0.62 &  <2.0  &  \cite{Petigura2017} \\ 
        K2-110 & K2V & K5 & 5.89 & 1.41$\pm$0.61 &  --  &  -- \\ 
        K2-111 & -- & G2 & 6.67 & 1.85$\pm$0.70 &  1.1$\pm$0.5  &  \cite{Mortier2020} \\ 
        K2-131 & -- & K5 & 7.77 & 3.80$\pm$0.45 &  4.3$\pm$0.2  &  \cite{Livingston2018} \\ 
        K2-136 & K5.5V & K5 & 6.63 & 2.35$\pm$0.52 &  3.0$\pm$0.5  &  \cite{Mann2018} \\ 
        K2-139 & K0V & G2 & 7.44 & 2.69$\pm$0.60 &  2.8$\pm$0.6  &  \cite{Barragan2018} \\ 
        K2-155 & K6V & K5 & 5.73 & 1.20$\pm$0.63 &  --  &  -- \\ 
        K2-167 & F7V & G2 & 7.61 & 2.88$\pm$0.58 &  --  &  -- \\ 
        K2-173 & -- & G2 & 6.42 & 1.58$\pm$0.74 &  --  &  -- \\ 
        K2-174 & K7V & K5 & 6.18 & 1.78$\pm$0.58 &  --  &  -- \\ 
        K2-179 & -- & G2 & 7.54 & 2.80$\pm$0.60 &  --  &  -- \\ 
        K2-180 & -- & K5 & 5.68 & 1.14$\pm$0.64 &  --  &  -- \\ 
        K2-182 & -- & G2 & 6.93 & 2.13$\pm$0.66 &  --  &  -- \\ 
        K2-188 & -- & G2 & 7.38 & 2.62$\pm$0.61 &  --  &  -- \\ 
        K2-189 & -- & G2 & 6.69 & 1.86$\pm$0.70 &  --  &  -- \\ 
        K2-198 & -- & K5 & 9.10 & 5.48$\pm$0.48 &  --  &  -- \\ 
        K2-199 & K5V & K5 & 6.19 & 1.80$\pm$0.57 &  --  &  -- \\ 
        K2-210 & -- & G2 & 6.67 & 1.85$\pm$0.70 &  --  &  -- \\ 
        K2-214 & -- & G2 & 7.36 & 2.60$\pm$0.61 &  --  &  -- \\ 
        K2-216 & K3V & K5 & 6.25 & 1.87$\pm$0.57 &  1.8$\pm$1.0  &  \cite{Persson2018} \\ 
        K2-222 & G0 & G2 & 7.37 & 2.62$\pm$0.61 &  --  &  -- \\ 
        K2-223 & -- & G2 & 7.03 & 2.24$\pm$0.65 &  --  &  -- \\ 
        K2-229 & -- & K5 & 6.63 & 2.35$\pm$0.52 &  2.46$\pm$0.22  &  \cite{Livingston2018} \\ 
        K2-232 & G0 & G2 & 7.58 & 2.84$\pm$0.59 &  4.16$\pm$0.282  &  \cite{Brahm2018} \\ 
        K2-261 & G7IV-V & G2 & 7.43 & 2.68$\pm$0.60 &  2.8$\pm$0.5  &  \cite{Johnson2018} \\ 
        K2-263 & G9V & K5 & 6.10 & 1.68$\pm$0.59 &  2.0$\pm$0.0  &  \cite{Mortier2018} \\ 
        K2-265 & ?+M4-7 & G2 & 6.89 & 2.09$\pm$0.67 &  --  &  -- \\ 
        K2-266 & K5V & K5 & 6.08 & 1.66$\pm$0.59 &  --  &  -- \\ 
        K2-280 & -- & G2 & 7.36 & 2.60$\pm$0.61 &  --  &  -- \\ 
        K2-285 & -- & K5 & 6.42 & 2.08$\pm$0.55 &  3.9$\pm$0.8  &  \cite{Palle2019} \\ 
        K2-290 & -- & G2 & 10.96 & 6.55$\pm$0.69 &  6.5$\pm$1.0  &  \cite{Hjorth2019} \\ 
        K2-291 & G0 & G2 & 7.10 & 2.32$\pm$0.64 &  2.0$\pm$0.0  &  \cite{Kosiarek2019b} \\ 
        KELT-4 A & -- & G2 & 10.30 & 5.81$\pm$0.61 &  6.0$\pm$1.2  &  \cite{Eastman2016} \\ 
        KELT-6 & F8 & G2 & 9.06 & 4.47$\pm$0.52 &  4.5$\pm$0.7  &  \cite{Biazzo2022} \\ 
        KELT-8 & G2V & G2 & 7.54 & 2.79$\pm$0.59 &  3.7$\pm$1.5  &  \cite{Bonomo2017} \\ 
        KELT-18 & F5 & G2 & 14.51 & 10.43$\pm$1.34 &  12.3$\pm$0.3  &  \cite{McLeod2017} \\ 
        KOI-94 & -- & G2 & 11.47 & 7.10$\pm$0.77 &  7.3$\pm$0.5  &  \cite{Weiss2013} \\ 
        KOI-351 & -- & G2 & 8.50 & 3.85$\pm$0.52 &  4.6$\pm$2.1  &  \cite{Cabrera2014} \\ 
        KOI-3680 & -- & G2 & 7.01 & 2.22$\pm$0.66 &  3.5$\pm$1.0  &  \cite{Hebrard2019} \\ 
        Kepler-9 & G2 & G2 & 7.52 & 2.78$\pm$0.59 &  2.74$\pm$0.4  &  \cite{Wang2018} \\ 
        Kepler-19 & -- & G2 & 6.71 & 1.89$\pm$0.70 &  2.0$\pm$0.0  &  \cite{Ballard2011} \\ 
        Kepler-20  & G5V & G2 & 6.88 & 2.08$\pm$0.67 &  2.0$\pm$0.0  &  \cite{Buchhave2016} \\ 
        Kepler-21  & F6IV & G2 & 11.50 & 7.13$\pm$0.78 &  8.4$\pm$0.5  &  \cite{Lopez2016} \\ 
        Kepler-25 & -- & G2 & 13.87 & 9.73$\pm$1.21 &  11.2$\pm$0.4  &  \cite{Steffen2012} \\ 
        Kepler-36 & -- & G2 & 8.76 & 4.13$\pm$0.52 &  4.9$\pm$1.0  &  \cite{Carter2012} \\ 
        Kepler-56 & -- & K5 & 6.82 & 2.58$\pm$0.50 &  --  &  -- \\ 
        Kepler-63 & G5 & G2 & 9.29 & 4.71$\pm$0.53 &  5.6$\pm$0.8  &  \cite{Sanchis2013} \\ 
        Kepler-65 & F6IV & G2 & 13.13 & 8.92$\pm$1.07 &  10.4$\pm$0.6  &  \cite{Chaplin2013} \\ 
        Kepler-68 & G1V & G2 & 7.15 & 2.37$\pm$0.64 &  0.5$\pm$0.5  &  \cite{Gilliland2013} \\ 
        Kepler-78 & K2 & G2 & 7.47 & 2.72$\pm$0.60 &  2.6$\pm$0.5  &  \cite{Grunblatt2015} \\ 
        Kepler-90 & -- & G2 & 8.50 & 3.85$\pm$0.52 &  --  &  -- \\ 
        Kepler-91 & K1IV & K5 & 7.09 & 2.94$\pm$0.48 &  3.2$\pm$0.5  &  \cite{Huber2013} \\ 
        Kepler-101 & -- & G2 & 7.34 & 2.58$\pm$0.61 &  2.6$\pm$0.5  &  \cite{Bonomo2014} \\ 
        Kepler-103 & G5Ib & G2 & 8.04 & 3.34$\pm$0.55 &  2.5$\pm$0.0  &  \cite{Marcy2014} \\ 
        Kepler-107 & -- & G2 & 7.80 & 3.09$\pm$0.56 &  3.6$\pm$0.5  &  \cite{Bonomo2019} \\ 
        Kepler-138 & M1V & K5 & 6.06 & 1.62$\pm$0.59 &  --  &  -- \\ 
        Kepler-381 & -- & G2 & 10.96 & 6.55$\pm$0.69 &  --  &  -- \\ 
        Kepler-421 & -- & G2 & 6.45 & 1.61$\pm$0.73 &  --  &  -- \\ 
        Kepler-423 & -- & G2 & 7.16 & 2.38$\pm$0.63 &  2.5$\pm$0.5  &  \cite{Gandolfi2015} \\ 
        Kepler-444 & K0V & K5 & 5.74 & 1.21$\pm$0.63 &  --  &  -- \\ 
        Kepler-449 & -- & G2 & 7.81 & 3.09$\pm$0.56 &  --  &  -- \\ 
        Kepler-489  & K2 & K5 & 6.08 & 1.65$\pm$0.59 &  --  &  -- \\ 
        Kepler-491 & K1 & K5 & 6.93 & 2.14$\pm$0.49 &  --  &  -- \\ 
        Kepler-495 & -- & G2 & 6.59 & 1.76$\pm$0.71 &  --  &  -- \\ 
        Kepler-511 & G0 & G2 & 7.48 & 2.73$\pm$0.60 &  --  &  -- \\ 
        Kepler-536 & -- & K5 & 6.33 & 1.97$\pm$0.56 &  --  &  -- \\ 
        Kepler-538 & G5V & G2 & 6.65 & 1.83$\pm$0.52 &  1.1$\pm$0.5  &  \cite{Mayo2019} \\ 
        Kepler-547 & -- & G2 & 6.96 & 2.16$\pm$0.66 &  --  &  -- \\ 
        Kepler-554 & -- & G2 & 6.94 & 2.15$\pm$0.66 &  --  &  -- \\ 
        Kepler-658 & M1V & G2 & 7.48 & 2.73$\pm$0.60 &  --  &  -- \\ 
        Kepler-849 & G0V & G2 & 9.17 & 4.58$\pm$0.52 &  --  &  -- \\ 
        Kepler-1323 & F8V & G2 & 8.33 & 3.67$\pm$0.53 &  --  &  -- \\ 
        Kepler-1514 & F7V & G2 & 12.10 & 7.79$\pm$0.88 &  --  &  -- \\ 
        Kepler-1515 & F6V & G2 & 15.56 & 11.57$\pm$1.56 &  --  &  -- \\ 
        Kepler-1655 & F5 & G2 & 7.87 & 3.15$\pm$0.56 &  3.5$\pm$0.5  &  \cite{Haywood2018} \\ 
        PH2 & -- & G2 & 6.94 & 2.14$\pm$0.66 &  1.43$\pm$0.78  &  \cite{Wang2013} \\ 
        Qatar-1 & -- & K5 & 6.58 & 2.28$\pm$0.53 &  1.9$\pm$0.7  &  \cite{Biazzo2022} \\
        Qatar-2 & K5V & K5 & 6.71 & 2.45$\pm$0.52 &  2.1$\pm$0.4  &  \cite{Biazzo2022} \\ 
        Ross 128 & dM4 & K5 & 5.01 & 0.29$\pm$0.74 &  --  &  -- \\ 
        TOI-561 & -- & G2 & 6.38 & 1.53$\pm$0.75 &  <2.0  &  \cite{Lacedelli2021} \\ 
        TYC 1422-614-1 & -- & K5 & 7.78 & 3.80$\pm$0.45 &  1.4$\pm$0.7  &  \cite{Niedzielski2015a} \\ 
        TYC 3318-01333-1 & -- & K5 & 7.32 & 3.22$\pm$0.47 &  1.5$\pm$0.7  &  \cite{Adamow2018} \\ 
        TYC 3667-1280-1 & -- & K5 & 9.28 & 5.72$\pm$0.50 &  3.2$\pm$0.4  &  \cite{Niedzielski2016b} \\ 
        TYC 4282-00605-1 & G/K & K5 & 7.38 & 3.30$\pm$0.46 &  3.0$\pm$0.5  &  \cite{Gonzalez2017} \\ 
        TrES-1 & -- & K5 & 6.27 & 1.90$\pm$0.56 &  1.3$\pm$0.3  &  \cite{Bonomo2017} \\ 
        TrES-2 & -- & G2 & 6.77 & 1.96$\pm$0.69 &  1.0$\pm$0.6  &  \cite{Winn2008} \\ 
        TrES-4 & -- & G2 & 13.41 & 9.23$\pm$1.12 &  9.5$\pm$0.8  &  \cite{Biazzo2022} \\ 
        WASP-1 & F7V & G2 & 7.80 & 3.08$\pm$0.56 &  5.77$\pm$0.35  &  \cite{Stempels2007} \\ 
        WASP-10 & K5V & K5 & 7.39 & 3.31$\pm$0.46 &  3.3$\pm$0.5  &  \cite{Biazzo2022} \\ 
        WASP-11 & K3V & K5 & 6.23 & 1.84$\pm$0.57 &  1.2$\pm$0.3  &  \cite{Biazzo2022} \\ 
        WASP-12 & G0V & G2 & 8.00 & 3.30$\pm$0.52 &  1.6$\pm$0.8  &  \cite{Albrecht2012} \\ 
        WASP-13 & G1V & G2 & 8.51 & 3.86$\pm$0.52 &  4.0$\pm$0.7  &  \cite{Biazzo2022} \\ 
        WASP-14 & F5V & G2 & 8.04 & 3.35$\pm$0.55 &  4.9$\pm$1.0  &  \cite{Bonomo2017} \\ 
        WASP-21 & G3V & G2 & 6.84 & 2.03$\pm$0.68 &  1.5$\pm$0.6  &  \cite{Bouchy2010} \\ 
        WASP-24 & F8-9 & G2 & 10.17 & 5.68$\pm$0.59 &  7.0$\pm$0.9  &  \cite{Bonomo2017} \\ 
        WASP-26 & G0 & G2 & 8.06 & 3.37$\pm$0.54 &  2.2$\pm$0.7  &  \cite{Albrecht2012} \\ 
        WASP-31 & F & G2 & 11.98 & 7.65$\pm$0.86 &  7.9$\pm$0.6  &  \cite{Anderson2011} \\ 
        WASP-32 & -- & G2 & 9.24 & 4.66$\pm$0.52 &  3.9$\pm$0.5  &  \cite{Bonomo2017} \\ 
        WASP-35 & -- & G2 & 7.88 & 3.17$\pm$0.56 &  3.9$\pm$0.4  &  \cite{Bonomo2017} \\ 
        WASP-36 & G2 & G2 & 7.60 & 2.86$\pm$0.58 &  3.3$\pm$1.2  &  \cite{Bonomo2017} \\ 
        WASP-38 & F8 & G2 & 12.23 & 7.94$\pm$0.90 &  8.3$\pm$0.8  &  \cite{Biazzo2022} \\ 
        WASP-39 & -- & K5 & 6.33 & 1.97$\pm$0.56 &  1.8$\pm$0.4  &  \cite{Biazzo2022} \\ 
        WASP-43 & K7V & K5 & 6.89 & 2.68$\pm$0.50 &  2.6$\pm$0.4  &  \cite{Biazzo2022} \\ 
        WASP-47 & -- & G2 & 7.25 & 2.48$\pm$0.62 &  1.8$\pm$0.24  &  \cite{Almenara2016} \\ 
        WASP-48 & -- & G2 & 17.09 & 13.26$\pm$1.87 &  12.2$\pm$0.7  &  \cite{Bonomo2017} \\
        WASP-49 & -- & G2 & 6.67 & 1.84$\pm$0.70 &  0.9$\pm$0.3  &  \cite{Bonomo2017} \\ 
        WASP-50 & G9V & G2 & 7.41 & 2.66$\pm$0.60 &  2.6$\pm$0.5  &  \cite{Bonomo2017} \\ 
        WASP-54 & F8 & G2 & 8.26 & 3.58$\pm$0.53 &  3.6$\pm$0.9  &  \cite{Biazzo2022} \\ 
        WASP-55 & G1 & G2 & 7.42 & 2.67$\pm$0.60 &  3.1$\pm$1.0  &  \cite{Bonomo2017} \\ 
        WASP-57 & G6 & G2 & 7.37 & 2.62$\pm$0.61 &  3.7$\pm$1.3  &  \cite{Bonomo2017} \\ 
        WASP-59 & K5V & K5 & 6.32 & 1.96$\pm$0.56 &  2.3$\pm$1.2  &  \cite{Hebrard2013} \\ 
        WASP-60 & G1V & G2 & 8.19 & 3.51$\pm$0.54 &  3.3$\pm$0.6  &  \cite{Biazzo2022} \\ 
        WASP-69 & -- & K5 & 6.76 & 2.51$\pm$0.51 &  2.2$\pm$0.4  &  \cite{Casasayas2017} \\ 
        WASP-74 & F9 & G2 & 9.32 & 4.75$\pm$0.53 &  6.03$\pm$0.19  &  \cite{Mancini2019} \\ 
        WASP-76 & F7 & G2 & 8.29 & 3.62$\pm$0.53 &  3.3$\pm$0.6  &  \cite{West2016} \\ 
        WASP-80 & K7V-M0V & K5 & 6.27 & 1.90$\pm$0.56 &  1.27$\pm$0.14  &  \cite{Triaud2015} \\ 
        WASP-84 & -- & K5 & 7.12 & 2.97$\pm$0.48 &  4.1$\pm$0.3  &  \cite{Anderson2014} \\ 
        WASP-85 A & -- & G2 & 7.47 & 2.72$\pm$0.60 &  --  &  -- \\ 
        WASP-103 & F8V & G2 & 14.79 & 10.73$\pm$1.40 &  10.6$\pm$0.9  &  \cite{Gillon2014} \\
        WASP-106 & F9 & G2 & 10.65 & 6.20$\pm$0.65 &  6.3$\pm$0.7  &  \cite{Bonomo2017} \\ 
        WASP-107 & K7V & K5 & 6.39 & 2.04$\pm$0.55 &  1.9$\pm$0.2  &  \cite{Mocnik2017b} \\ 
        WASP-118 & -- & G2 & 14.72 & 10.65$\pm$1.38 &  9.68$\pm$1.14  &  \cite{Hay2016} \\
        WASP-127 & G5 & G2 & 6.94 & 2.14$\pm$0.66 &  0.3$\pm$0.2  &  \cite{Lam2017} \\ 
        WASP-135 & G5V & G2 & 7.58 & 2.84$\pm$0.59 &  4.67$\pm$0.89  &  \cite{Spake2016} \\ 
        WASP-148 & -- & G2 & 7.20 & 2.43$\pm$0.63 &  3.0$\pm$2.0  &  \cite{Hebrard2020} \\ 
        WASP-162 & K0 & G2 & 7.49 & 2.74$\pm$0.60 &  1.0$\pm$0.8  &  \cite{Hellier2019} \\ 
        XO-1 & G1V & G2 & 6.99 & 2.19$\pm$0.66 &  1.11$\pm$0.067  &  \cite{McCullough2006} \\ 
        XO-2 N & G9V & K5 & 6.56 & 2.27$\pm$0.53 &  1.8$\pm$0.4  &  \cite{Biazzo2022} \\ 
        XO-2 S & G9V & K5 & 6.47 & 2.15$\pm$0.54 &  1.7$\pm$0.4  &  \cite{Biazzo2022} \\ 
        XO-4 & -- & G2 & 12.33 & 8.04$\pm$0.92 &  8.8$\pm$0.5  &  \cite{Bonomo2017} \\ 
        XO-5 & G8V & G2 & 7.01 & 2.22$\pm$0.65 &  0.7$\pm$0.5  &  \cite{Bonomo2017} \\ 
        $\kappa$ CrB & K1IVa & K5 & 6.76 & 2.52$\pm$0.51 &  1.5$\pm$0.5  &  \cite{Johnson2008} \\ 
        $\tau$ Cet & G8V & G2 & 6.23 & 1.37$\pm$0.77 &  --  &  -- \\
        \hline
\end{longtable}
}

\begin{figure}
    \centering
    \includegraphics[width=\columnwidth]{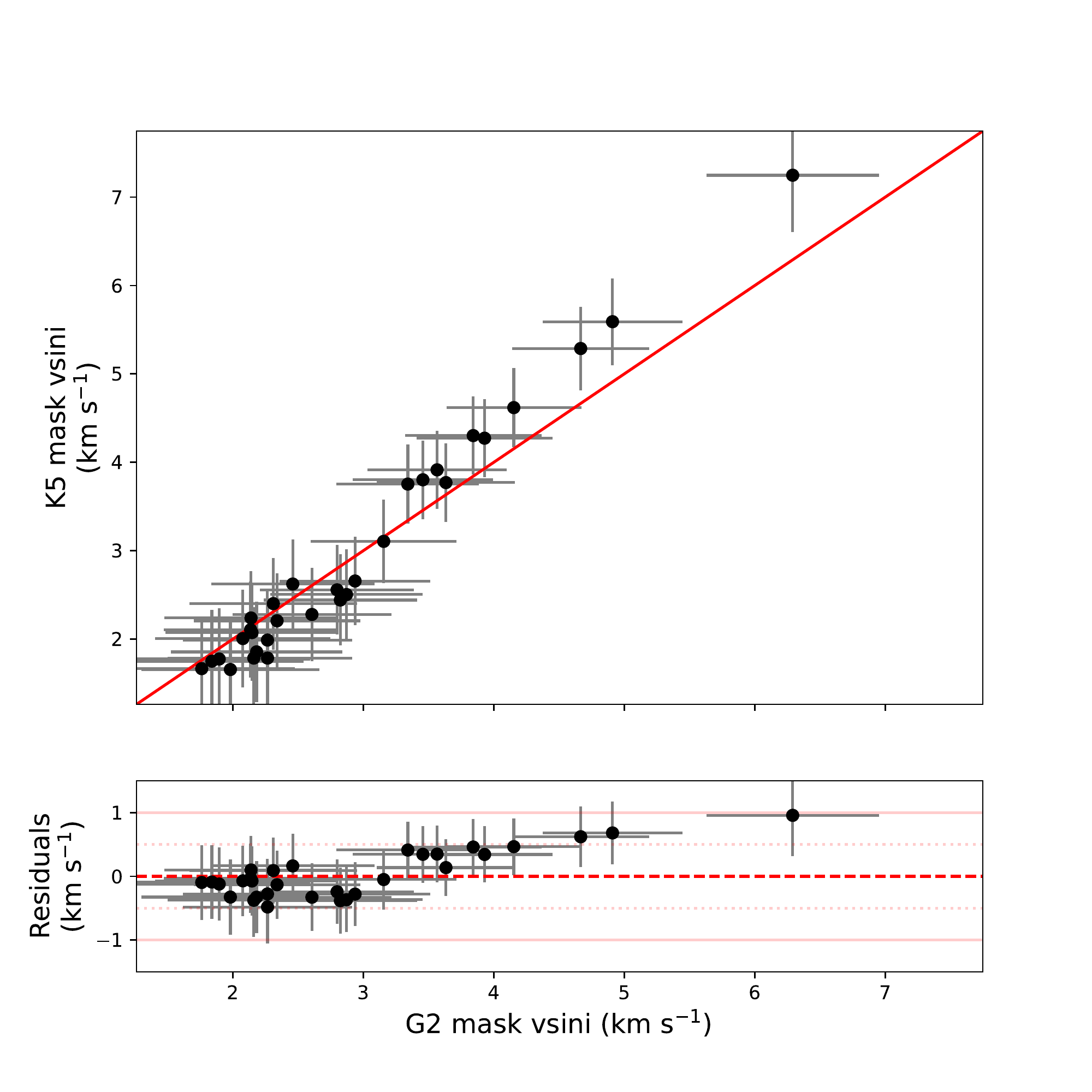}
    \caption{\textit{Upper panel:} comparison between $v_{\mathrm{eq}}\sin{i_\star}$ obtained with the G2 relation ($x$-axis) and the K5 relation ($y$-axis). The red line shows the one-to-one correlation. \textit{Lower panel:} residuals.}
    \label{fig:g2_k5_comparison}
\end{figure}

Some of the objects in our sample have both G2 and K5 CCFs in the TNG archive, and so we were able to directly compare the results of the two calibrations, in order to quantify the effect of a spectral type mismatch on the resulting $v_{\mathrm{eq}}\sin{i_\star}$ (see Fig.~\ref{fig:g2_k5_comparison}). These objects have a relatively small range of $v_{\mathrm{eq}}\sin{i_\star}$, but still the results agree with less than 0.5 km~s$^{-1}$ difference for $v_{\mathrm{eq}}\sin{i_\star}$ < 4 km~s$^{-1}$, and with less than 1 km~s$^{-1}$ for $v_{\mathrm{eq}}\sin{i_\star}$ > 4 km~s$^{-1}$.
Still, to ensure the best possible result, care should be taken to reduce every star with the more appropriate mask. Usually this is already done, because the better the star-mask match, the smaller is the error of the radial velocity computed by the DRS, but sometimes the stellar type is unknown prior to the observations and a mismatch may occur. Possible mismatches between hotter stars (early F-type or above) and the G2 mask are not considered here because hotter stars are usually also fast rotators and they would naturally fall outside the applicability range of our relation (FWHM$_\mathrm{DRS}$ < 20 km~s$^{-1}$). Because we relied on the public data present in the TNG archive, there are a few mismatches between stellar type and mask in our sample, but in all these cases we have $v_{\mathrm{eq}}\sin{i_\star} <$ 4 km~s$^{-1}$, so the mismatches should not affect heavily the results.

\subsection{Comparison with the literature}
\label{subsec:vsini_literature}
\begin{figure}
    \centering
    \includegraphics[width=\columnwidth]{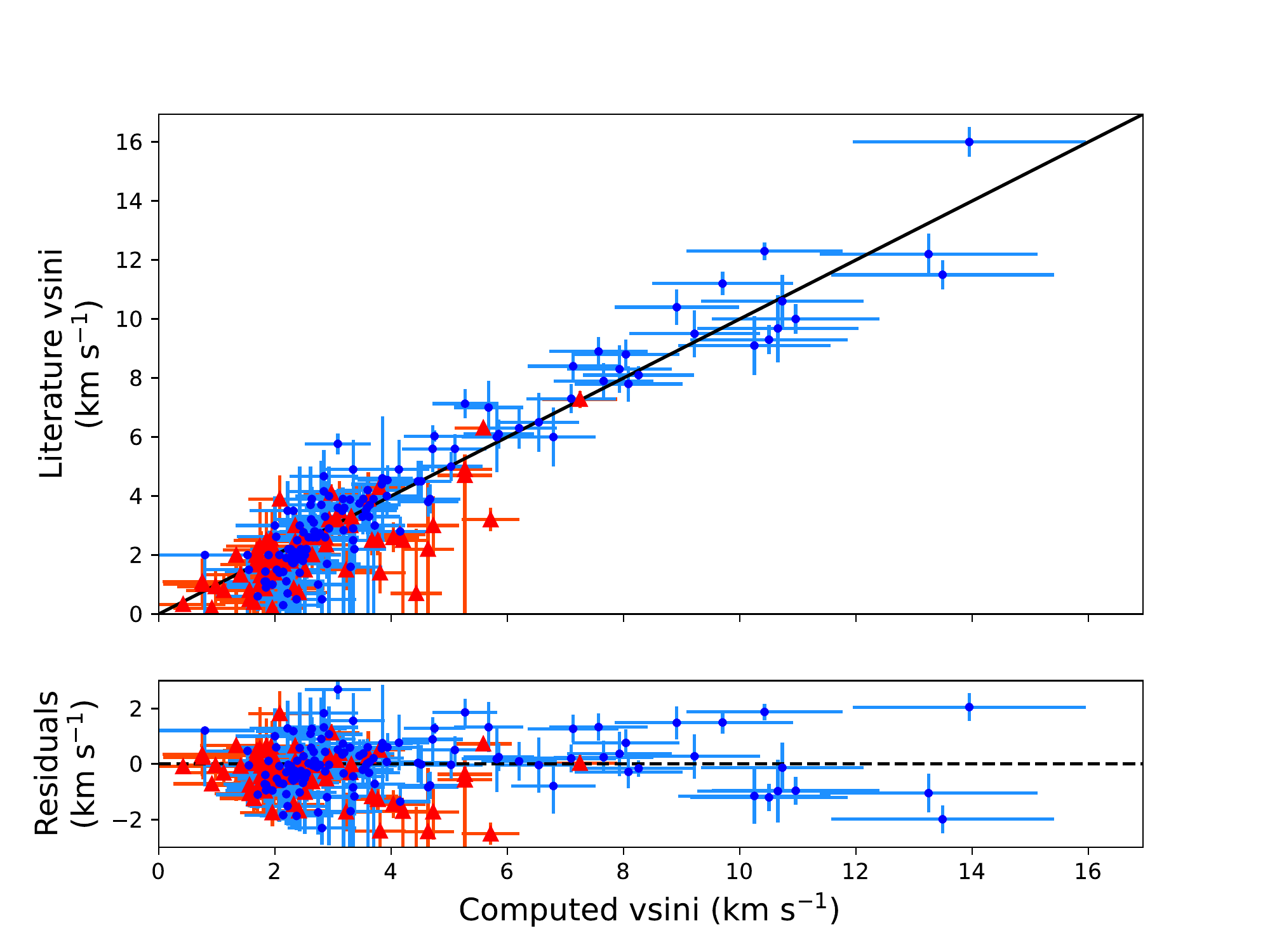}
    \caption{\textit{Upper panel:} comparison between $v_{\mathrm{eq}}\sin{i_\star}$ values from literature (\textit{y}-axis) and estimated from the CCF FWHM$_\mathrm{DRS}$ (\textit{x}-axis). Blue dots are values computed with the G2 mask relation, red triangles are those computed with the K5 mask relation. The black line shows the one-to-one correlation. \textit{Lower panel:} residuals of the one-to-one correlation shown above.}
    \label{fig:comparison}
\end{figure}

Out of the stars listed in Table~\ref{tab:all_vsini}, 206 had also $v_{\mathrm{eq}}\sin{i_\star}$ values from the literature, so we could compare our results with them (see Fig.~\ref{fig:comparison}).
As a sanity check, we used this larger sample to test our relations: we calibrated the G2 and K5 FWHM$_\mathrm{DRS}$ values using the the whole set of literature $v_{\mathrm{eq}}\sin{i_\star}$ values. The resulting relations are:

\begin{equation}
\label{eq:vsini_literature}
    \begin{split}
        \mathrm{G2\ mask:\ } v_{\mathrm{eq}}\sin{i_\star} = 1.1241 \times \mathrm{FWHM}_\mathrm{DRS} - 5.70685~~ \\
        \mathrm{K5\ mask:\ } v_{\mathrm{eq}}\sin{i_\star} = 0.95935 \times \mathrm{FWHM}_\mathrm{DRS} - 4.37978
    \end{split}
\end{equation}

\begin{figure}
    \centering
    \includegraphics[width=\columnwidth]{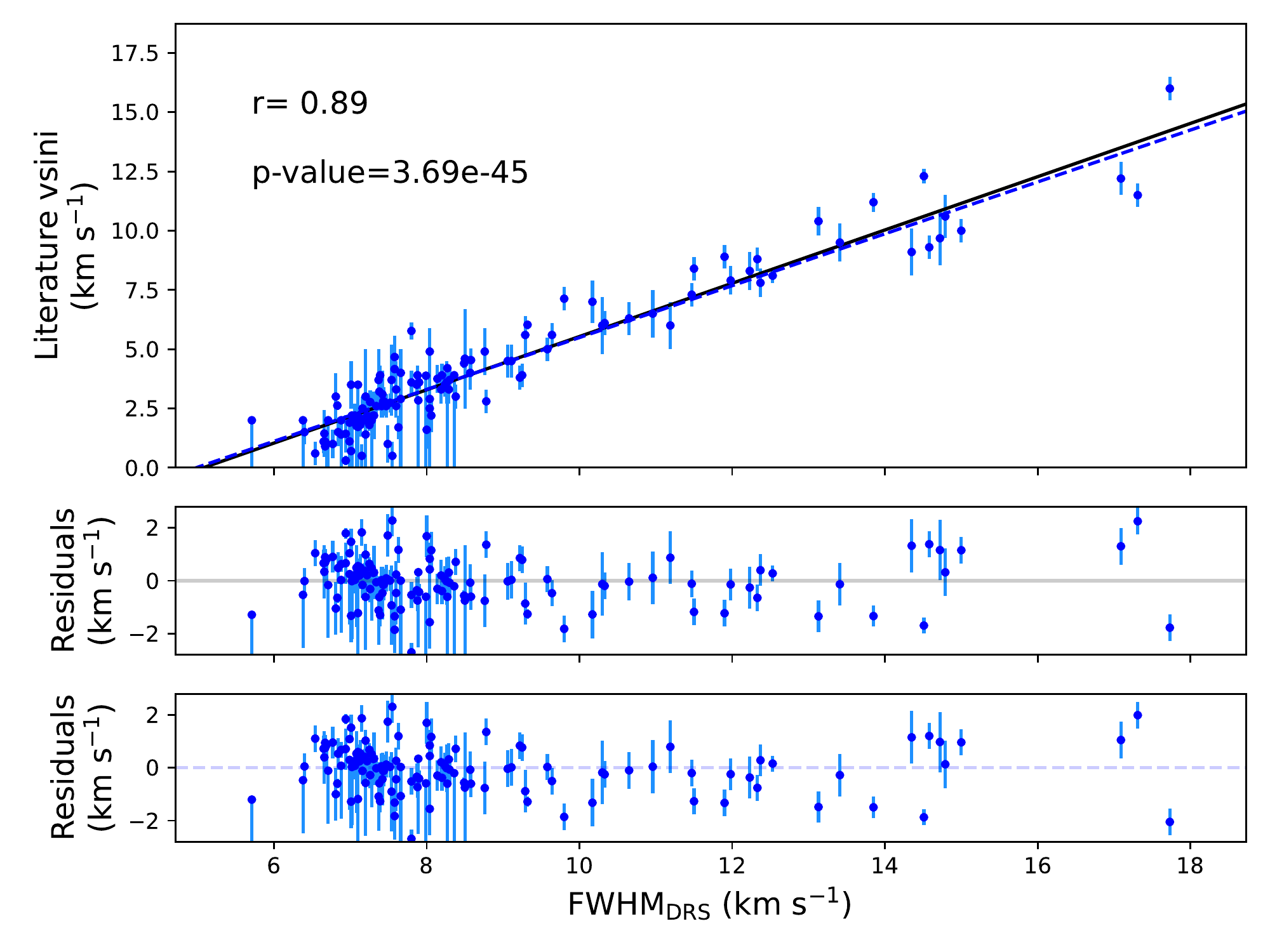}
    \caption{\textit{Upper panel:} correlation between the G2 FWHM$_\mathrm{DRS}$ ($x$-axis) and the literature $v_{\mathrm{eq}}\sin{i_\star}$ values ($y$-axis), with the Spearman's correlation coefficient $r$ and $p$-value shown in the plot. The black line shows the linear fit of the data, the blue dashed line shows the relation obtained from our selected calibrators (Eq.~\ref{eq:vsini}). \textit{Middle panel:} residuals of the linear fitting. \textit{Lower panel:} residuals of the relation from selected calibrators.}
    \label{fig:g2_literature} 
\end{figure}

\begin{figure}
    \includegraphics[width=\columnwidth]{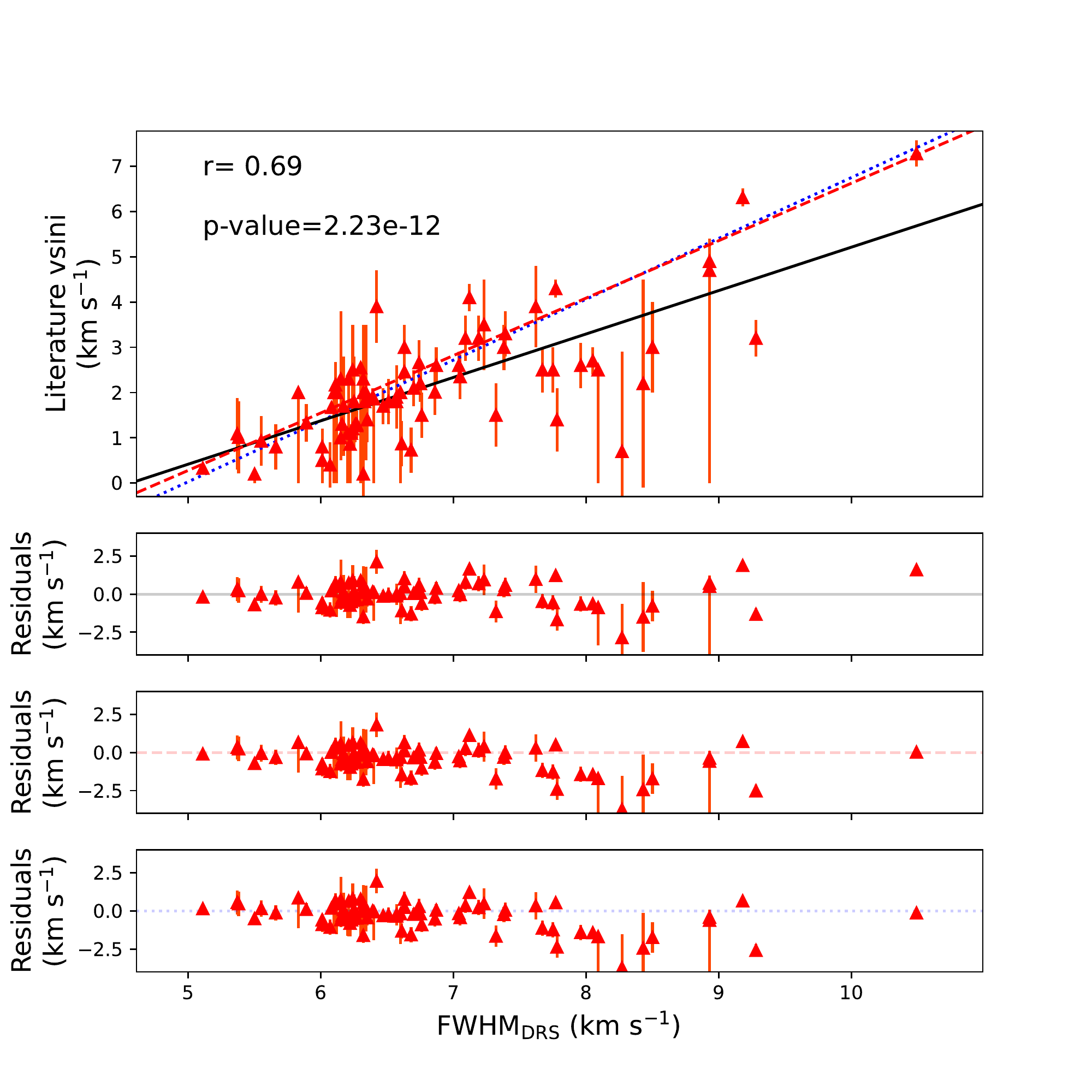}
    \caption{\textit{Upper panel:} correlation between the K5 FWHM$_\mathrm{DRS}$ ($x$-axis) and the literature $v_{\mathrm{eq}}\sin{i_\star}$ values ($y$-axis), with the Spearman's correlation coefficient $r$ and $p$-value shown in the plot. The black line shows the linear fit of the data, the red dashed line shows the relation obtained from our selected calibrators (Eq.~\ref{eq:vsini}), the blue dotted line shows the linear fit after removing the stars with $\log g < 3.5$. \textit{Lower panels:} residuals of the linear fitting, of the relation from selected calibrators, and of the linear fitting after removing the stars with $\log g < 3.5$, respectively.}
    \label{fig:k5_literature}
\end{figure}

\begin{figure}
    \centering
    \includegraphics[width=\columnwidth]{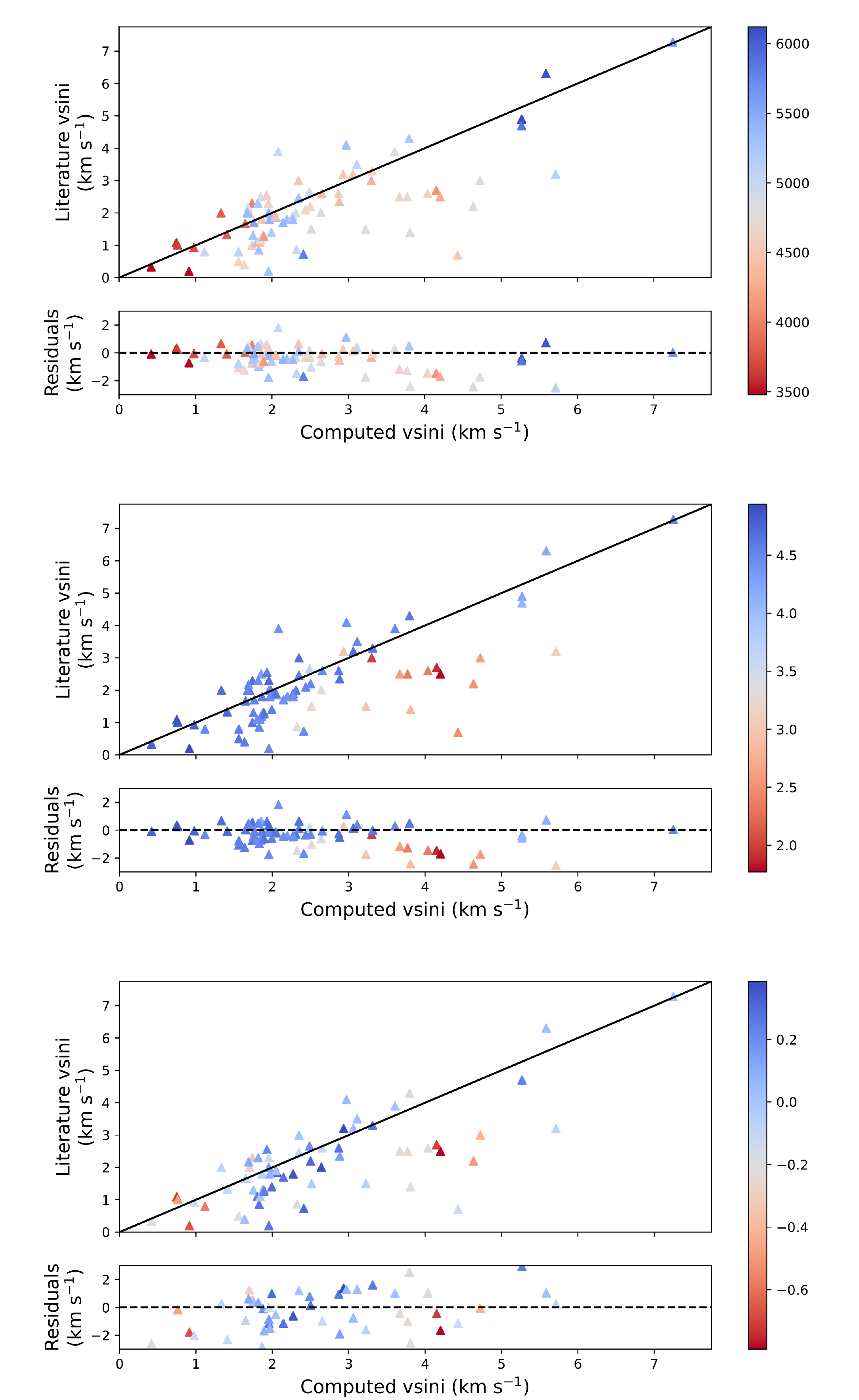}
    \caption{Comparison between our $v_{\mathrm{eq}}\sin{i_\star}$ ($x$-axis) and the literature values ($y$-axis) when using the K5 relation, color-coded according to the stellar parameters $T_{\mathrm{eff}}$ (\textit{upper panel}), $\log g$ (\textit{middle panel}), and [Fe/H] (\textit{lower panel}).}
    \label{fig:k5_parameters}
\end{figure}

As it is shown in Fig.~\ref{fig:g2_literature}, there is almost no difference between the relation obtained using the whole literature data set and the original one obtained from the selected calibrators (Table~\ref{tab:calibrators}) for the G2 mask, while the situation is different when using the K5 mask (see black solid line and red dashed line in Fig.~\ref{fig:k5_literature}). In this case, the spread is larger (and the Spearman's $r$ coefficient lower), and so is the difference between the original calibration and the new one. We also lack reliable data points with FWHM$_\mathrm{DRS}$ > 12 km~s$^{-1}$, and the literature $v_{\mathrm{eq}}\sin{i_\star}$ values are very spread out. The latter fact could be caused by the type of stars that are usually reduced using the K5 mask, i.e. mid and late K-type and early M-type stars: these objects may be very active and this could affect both the shape of the CCF (and thus the FWHM$_\mathrm{DRS}$) and the $v_{\mathrm{eq}}\sin{i_\star}$ estimation performed in literature.
To better investigate this behaviour, and to check the possible limitations of our relations' applicability range, we looked at the sample considering also the stellar parameters of the stars, i.e. $T_{\mathrm{eff}}$, $\log g$, and [Fe/H]. We recovered the parameters from SIMBAD\footnote{\url{http://simbad.u-strasbg.fr/simbad/}} \citep{Wenger2000} using an automated python query. We show the results in Fig.~\ref{fig:g2_parameters} for the G2 relation, and in Fig.~\ref{fig:k5_parameters} for the K5 relation. While there is no obvious trend looking at the results from the G2 relation, we can see that stars with $\log g < 3.5$ tend to cluster below the one-to-one correlation when comparing the results from the K5 relation to the literature $v_{\mathrm{eq}}\sin{i_\star}$ values. If we perform a linear fit between our $v_{\mathrm{eq}}\sin{i_\star}$ and the literature $v_{\mathrm{eq}}\sin{i_\star}$ only for stars with $\log g > 3.5$ (blue dotted line in Fig.~\ref{fig:k5_literature}), then the resulting relation agrees much better with that obtained from the selected calibrators:
\begin{figure}
    \centering
    \includegraphics[width=\columnwidth]{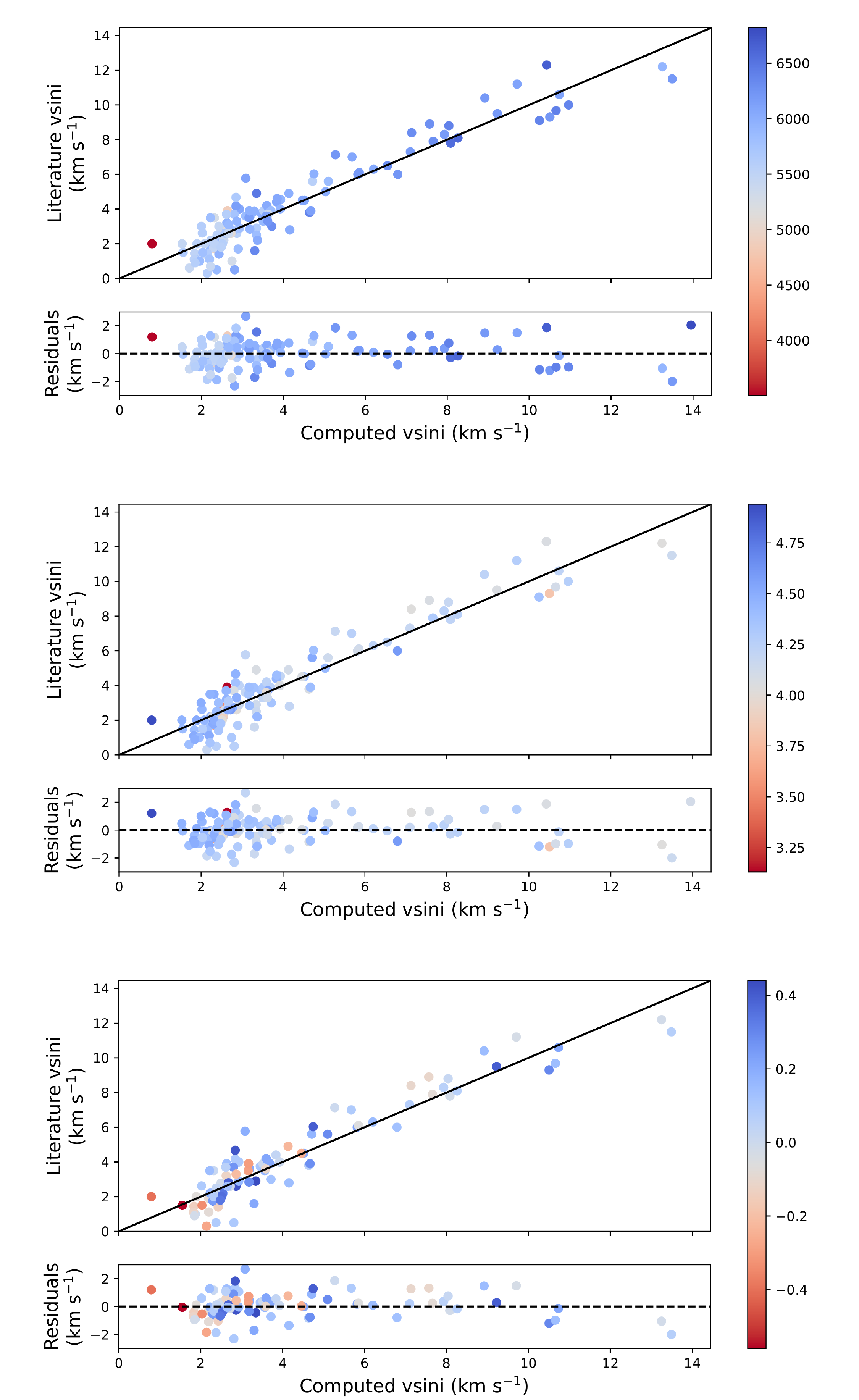}
    \caption{Comparison between our $v_{\mathrm{eq}}\sin{i_\star}$ ($x$-axis) and the literature values ($y$-axis) when using the G2 relation, color-coded according to the stellar parameters $T_{\mathrm{eff}}$ (\textit{upper panel}), $\log g$ (\textit{middle panel}), and [Fe/H] (\textit{lower panel}).}
    \label{fig:g2_parameters}
\end{figure}
\begin{equation}
    \label{eq:vsini_lit_k5}
    \mathrm{K5\ mask:\ } v_{\mathrm{eq}}\sin{i_\star} = 1.34470 \times \mathrm{FWHM}_\mathrm{DRS} - 6.69438
\end{equation}
While we advise using Eq.~\ref{eq:vsini} to compute $v_{\mathrm{eq}}\sin{i_\star}$ because we better trust our selected calibrators, we list in Table~\ref{tab:errors} also the parameters' errors and correlation factors needed to compute the errors when using Eq.~\ref{eq:vsini_literature} (G2 mask only), and Eq.~\ref{eq:vsini_lit_k5} (K5 mask).

We can assume that, at least in the case of the K5 sample, our relations are applicable only for stars with $\log g > 3.5$, i.e. mostly main sequence stars, but also some subgiants and red giants stars may fall in the applicability range. Unfortunately, we do not have a wide enough range of $\log g$ values in our G2 sample to test the same behaviour (see Fig.~\ref{fig:g2_parameters}, middle panel), but considering that the G2 mask used in the HARPS-N DRS is optimized for the Sun, we can infer that also the G2 relation is best suited for main-sequence stars.

Comparing our results with the literature $v_{\mathrm{eq}}\sin{i_\star}$, we found no star where our $v_{\mathrm{eq}}\sin{i_\star}$ differs more the 3$\sigma$ from the literature value, and only 4 where the difference is larger than 2$\sigma$ (WASP-1, WASP-127, TYC 1422-614-1, and TYC 3667-1280-1). Taking into account the very different methods used in literature to compute $v_{\mathrm{eq}}\sin{i_\star}$ this is a good indicator of the robustness and reliability of our FWHM$_\mathrm{DRS}$-$v_{\mathrm{eq}}\sin{i_\star}$ relation.

\subsection{Stellar inclination}
\label{subsec:characterization}
We focused on the results we obtained for stars with no $v_{\mathrm{eq}}\sin{i_\star}$ literature value, to see if we were able to recover an estimate of the stellar inclination $i_\star$. We did not perform this work on the other targets because our results do not differ much from those already in the literature, and so we do not expect any substantial changes or improvements on $i_\star$.

We used Eq.~\ref{eq:inclination} to compute $i_\star$, which means that we could work only with objects with known $P_\mathrm{rot}$ and $R_\star$. In some cases, the exoplanetary orbit inclination was known: we could then compare it to $i_\star$, to check the spin-orbit alignment of the system. Because of the sometimes large errors on the various parameters, many $i_\star$ results were compatible with the whole range of possible inclinations.

We show in Table~\ref{tab:inclination} only the results that set some constrains on the stellar possible inclination.
\begin{table*}[]
    \centering
    \caption{Stellar inclination $i_\star$ derived from our $v_{\mathrm{eq}}\sin{i_\star}$ values, compared with the planetary orbit inclination $i_p$, if known.}
    \begin{tabular}{c|c|c|c|c|cc}
    \hline\hline
        Name & $P_\mathrm{rot}$ & $R_\star$ & $v_{\mathrm{eq}}\sin{i_\star}$ & $i_\star$ & \multicolumn{2}{c}{$i_p$} \\
        & [days] & [$R_\sun$] & [km~s$^{-1}$] & [deg] & \multicolumn{2}{c}{[deg]} \\
         \hline
        GJ 328 & 33.6$^a$ & 0.65$\pm$0.02$^a$ & 1.55$\pm$0.60$^b$ & > 75$^b$ & \multicolumn{2}{c}{--} \\
        HD 13931 & 26$^c$ & 1.23$\pm$0.06$^c$ & 2.41$\pm$0.63$^b$ & > 47$^b$ & \multicolumn{2}{c}{39$\substack{+13 \\ -8}$ planet b$^d$} \\
        HD 26965 & 37--43$^e$ & 0.87$\pm$0.17$^f$ & 1.44$\pm$0.61$^b$ & > 40$^b$ & \multicolumn{2}{c}{--} \\
        K2-3 & 40$\pm$2$^g$ & 0.561$\pm$0.068$^h$ & 1.21$\pm$0.63$^b$ & > 52$^b$ & 89.588$\substack{+0.116 \\ -0.100}$ & planet b$^g$ \\
        K2-79 & 29.08$\pm$6.20$^i$ & 1.247$\substack{+0.077 \\ -0.072}^l$ & 2.46$\pm$0.63$^b$ & > 48$^b$ & 88.63$\substack{+0.98 \\ -1.66}^l$ & planet b \\
        K2-155 & 34.8$\pm$8.2$^l$ & 0.58$\substack{+0.06 \\ -0.03}^m$ & 1.20$\pm$0.63 $^b$ & > 36$^b$ & 88.3$\substack{+1.2 \\ -1.9}$ & planet b$^m$ \\
        & & & & & 88.96$\substack{+0.71 \\ -0.88}$ & planet c$^m$ \\
        & & & & & 89.61$\substack{+0.27 \\ -0.48}$ & planet d$^m$ \\
        K2-173 & 20.31$\pm$2.12$^i$ & 1.00$\pm$0.08$^n$ & 1.58$\pm$0.74$^b$ & 39$\pm$23$^b$ & $87.83\substack{+1.54 \\ -2.87}$  & planet b$^i$\\
        K2-198 & 6.97$\pm$0.41$^i$ & $0.78\substack{+0.03 \\ - 0.05}$$^n$ & 5.48$\pm$0.48$^b$ & 75$\substack{+15 \\ -27}^b$ & 88.904$\substack{+0.094 \\-0.027}$  & planet b$^o$ \\
         & & & & & 86.494$\substack{+0.268 \\ -0.088}$ & planet c$^o$ \\
          & & & & & 86.494$\substack{+0.268 \\ -0.088}$ & planet d$^o$ \\
        Kepler-495 & 19.20$\pm$2.98$^p$ & 0.867$\substack{+0.039 \\ -0.037}^q$ & 1.76$\pm$0.71$^b$ & 50$\pm$30$^b$ & \multicolumn{2}{c}{--} \\
        Kepler-849 & 17.91$\pm$0.48$^p$ & 1.828$\substack{+0.086 \\ -0.081}^q$ & 4.58$\pm$0.52$^b$ & 62$\pm$14$^b$ & \multicolumn{2}{c}{--} \\
        Kepler-1514 & 7.83$\pm$0.16$^p$ & 1.273$\substack{+0.055 \\ -0.052}^q$ & 7.79$\pm$0.88$^b$ & 72$\substack{+18 \\ -21}^b$ &  89.944$\substack{+0.013 \\ -0.010}$ & planet b$^r$\\
        & & & & & 87.98$\substack{+1.20 \\ -0.40}$ & planet c$^r$ \\
        WASP-85 A & 15.1$\pm$0.6$^s$ & 0.935$\pm$0.023$^s$ & 2.72$\pm$0.60$^b$ & 60$\pm$22$^b$ & 89.69$\substack{+0.11 \\ -0.03}$ & planet b$^s$ \\       
    \hline
    \end{tabular}
    \tablebib{$^a$~\citet{Kuker2019}; $^b$~this work; $^c$~\citet{Howard2010}; $^d$~\cite{Philipot2023};
    $^e$~\citet{Diaz2018}; $^f$~\citet{Ma2018}; $^g$~\citet{Kosiarek2019}; $^h$~\citet{Crossfield2015}; $^i$~\citet{Reinhold2020};$^l$~\citet{Mayo2018}; $^m$~\citet{DiezAlonso2018};    $^n$~\citet{Stassun2019};  $^o$~\citet{Hedges2019}; $^p$~\citet{Mazeh2015}; $^q$~\citet{Berger2018}; $^r$~\citet{Dalba2021}; $^s$~\citet{Mocnik2016}}
    \label{tab:inclination}
\end{table*}
While in most cases our results are compatible with aligned, edge-on planetary systems, we still found one system that shows a difference between $i_\star$ and $i_p$ around the 2$\sigma$ level (K2-173), and hints that the HD 13931 system may be aligned, but not edge-on.

\section{Extension to other spectrographs}
\label{sec:other_spectrographs}

The relations found in our work between FWHM$_\mathrm{DRS}$ and $v_{\mathrm{eq}}\sin{i_\star}$ are optimized for a specific combination of instrument, software and stellar masks. While there are other spectrographs with dedicated DRS, and a few of them also deliver the spectra's CCFs as an output, the different resolution, instrumental effects, wavelength ranges, numerical codes used to compute the CCF, and stellar masks could heavily influence the FWHM$_\mathrm{DRS}$--$v_{\mathrm{eq}}\sin{i_\star}$ relation. A possible exception could be the HARPS spectrograph \citep{Mayor2003}, of which HARPS-N is a twin, not only concerning the hardware, but also the software, as HARPS and HARPS-N have almost the same DRS.

To test this assumption, we checked the public archives of two spectrographs with a similar spectral range as HARPS-N: HARPS (which has also the same resolution, telescope aperture and DRS as HARPS-N) and SOPHIE\footnote{\url{http://www.obs-hp.fr/guide/sophie/sophie-eng.shtml}}. Both spectrographs have been used for many years in the exoplanets' search and characterization field, guaranteeing the availability of a large amount of public data of exoplanet-host stars. The main characteristics of HARPS-N, HARPS, and SOPHIE are listed in Table~\ref{tab:spectrographs}. SOPHIE has an high-resolution (HR) and an high-efficiency (HE) mode, but for a more direct comparison with HARPS-N we focused on the HR mode spectra to start.
\begin{table}[]
    \centering
    \caption{Main characteristics of the HARPS-N, HARPS, and SOPHIE spectrographs.}
    \begin{tabular}{c|c|c|c}
    \hline\hline
    Spectrograph & Telescope & Wavelength & Resolution \\
     & diameter [m] & range [nm] & \\
    \hline
       HARPS-N  & 3.58 & 385--691 & 115,000 \\
       HARPS & 3.57 & 378--691 & 115,000 \\
       SOPHIE (HR) & 1.93 & 387--694 & 75,000 \\
       SOPHIE (HE) & 1.93 & 387--694 & 40,000 \\
    \hline
    \end{tabular}
    \label{tab:spectrographs}
\end{table}
Both HARPS and SOPHIE have dedicated DRS that deliver the spectra's CCFs and their FWHMs using stellar masks similar (or, in the case of HARPS, identical) to the HARPS-N ones. We note here that also SOPHIE DRS is adapted from the HARPS DRS, so the three instruments have the same or a very similar DRS.

\begin{figure}
    \centering
    \includegraphics[width=\columnwidth]{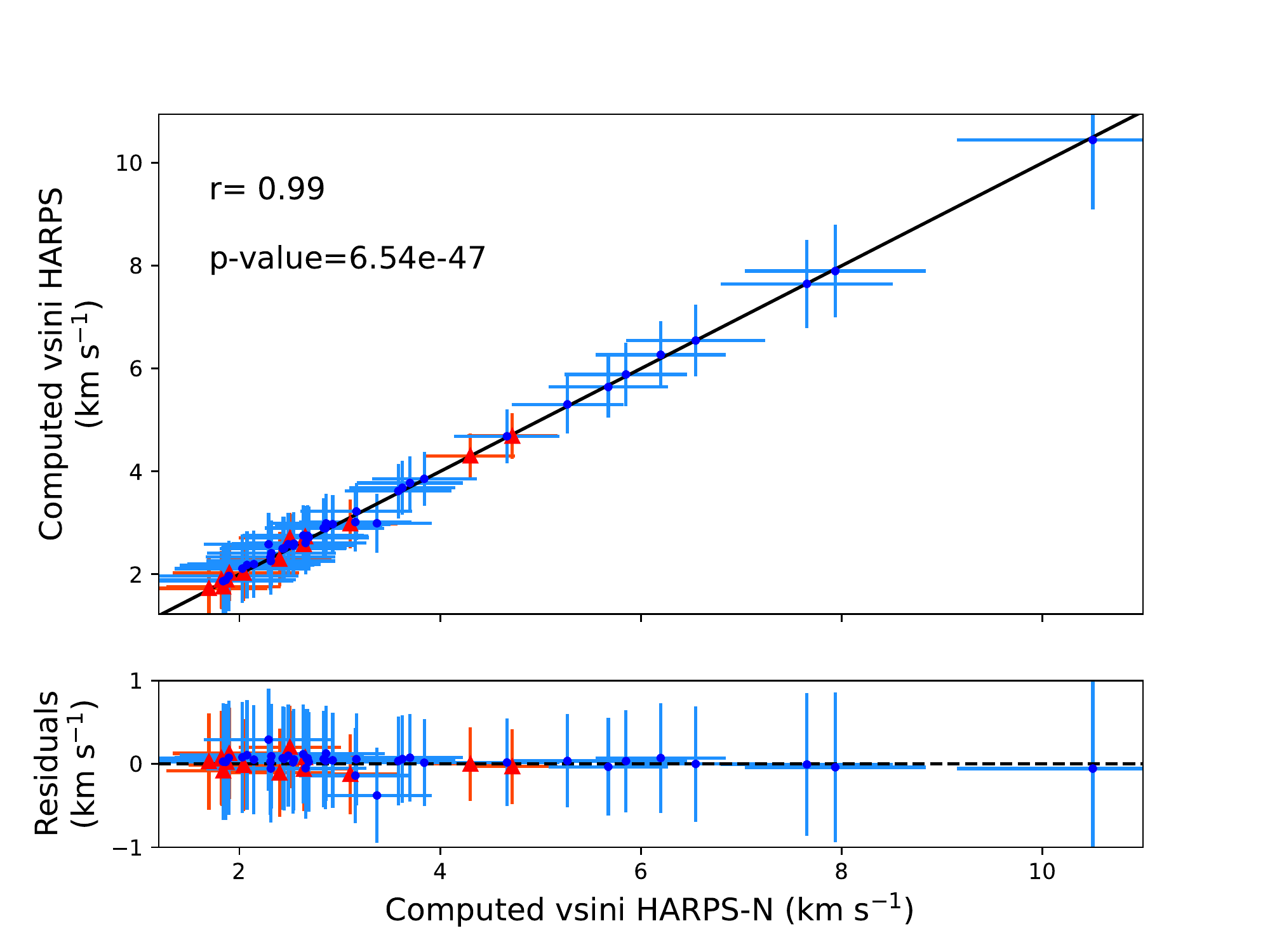}
    \caption{\textit{Upper panel:} comparison between $v_{\mathrm{eq}}\sin{i_\star}$ computed from the HARPS-N FWHM$_\mathrm{DRS}$ ($x$-axis) and those computed from the HARPS FWHM$_\mathrm{DRS}$ ($y$-axis). The blue dots are the values computed with the G2 mask relation, the red triangles are those computed with the K5 mask relation. The black line shows the one-to-one correlation. \textit{Lower panel:} residuals. }
    \label{fig:comparisonHARPS}
\end{figure}

\begin{figure}
    \centering    \includegraphics[width=\columnwidth]{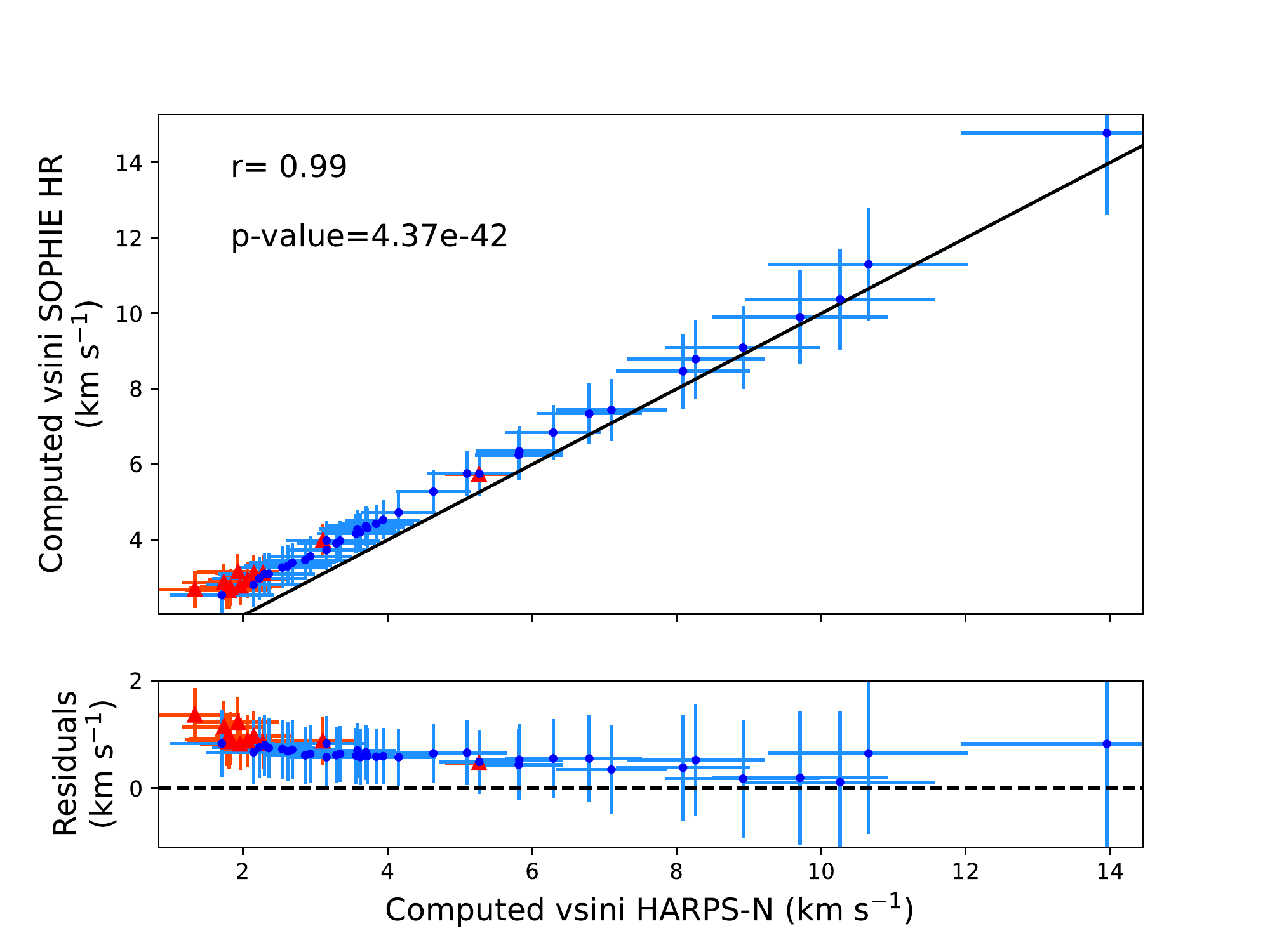}
    \caption{\textit{Upper panel:} comparison between $v_{\mathrm{eq}}\sin{i_\star}$ computed from the HARPS-N FWHM$_\mathrm{DRS}$ ($x$-axis) and those computed from the SOPHIE high-resolution FWHM$_\mathrm{DRS}$ ($y$-axis). The blue dots are the values computed with the G2 mask relation, the red triangles are those computed with the K5 mask relation. The black line shows the one-to-one correlation. \textit{Lower panel:} residuals.}
    \label{fig:comparisonSOPHIE}
\end{figure}

We searched the dedicated HARPS\footnote{\url{http://archive.eso.org/scienceportal/home}} and SOPHIE\footnote{\url{http://atlas.obs-hp.fr/sophie/}} archives for objects listed in Table~\ref{tab:all_vsini} to download their HARPS and SOPHIE CCFs. We selected only the CCFs obtained with either the G2 or K5 mask in the high-resolution mode, up to a maximum of 50 per object, so that, when possible, we could recover a statistically robust median FWHM$_\mathrm{DRS}$ for each object. We then computed the $v_{\mathrm{eq}}\sin{i_\star}$ from the median FWHM$_\mathrm{DRS}$ using Eq.~\ref{eq:vsini}, and we compared the results with our HARPS-N $v_{\mathrm{eq}}\sin{i_\star}$. Figure~\ref{fig:comparisonHARPS} shows the comparison between the HARPS-N and HARPS results, and Fig.~\ref{fig:comparisonSOPHIE} shows the comparison between the HARPS-N and SOPHIE results.

It is plainly visible that the twin status of the HARPS and HARPS-N spectrographs would allow us to use the HARPS-N calibration directly with the HARPS data. It is interesting to note that, because we used HARPS spectra observed both before and after 2015, this is true for HARPS data taken both before and after the change of fibers \citep{LoCurto2015}, even if this change should have slightly affected the FWHM$_\mathrm{DRS}$.

\begin{figure}
    \centering
    \includegraphics[width=\columnwidth]{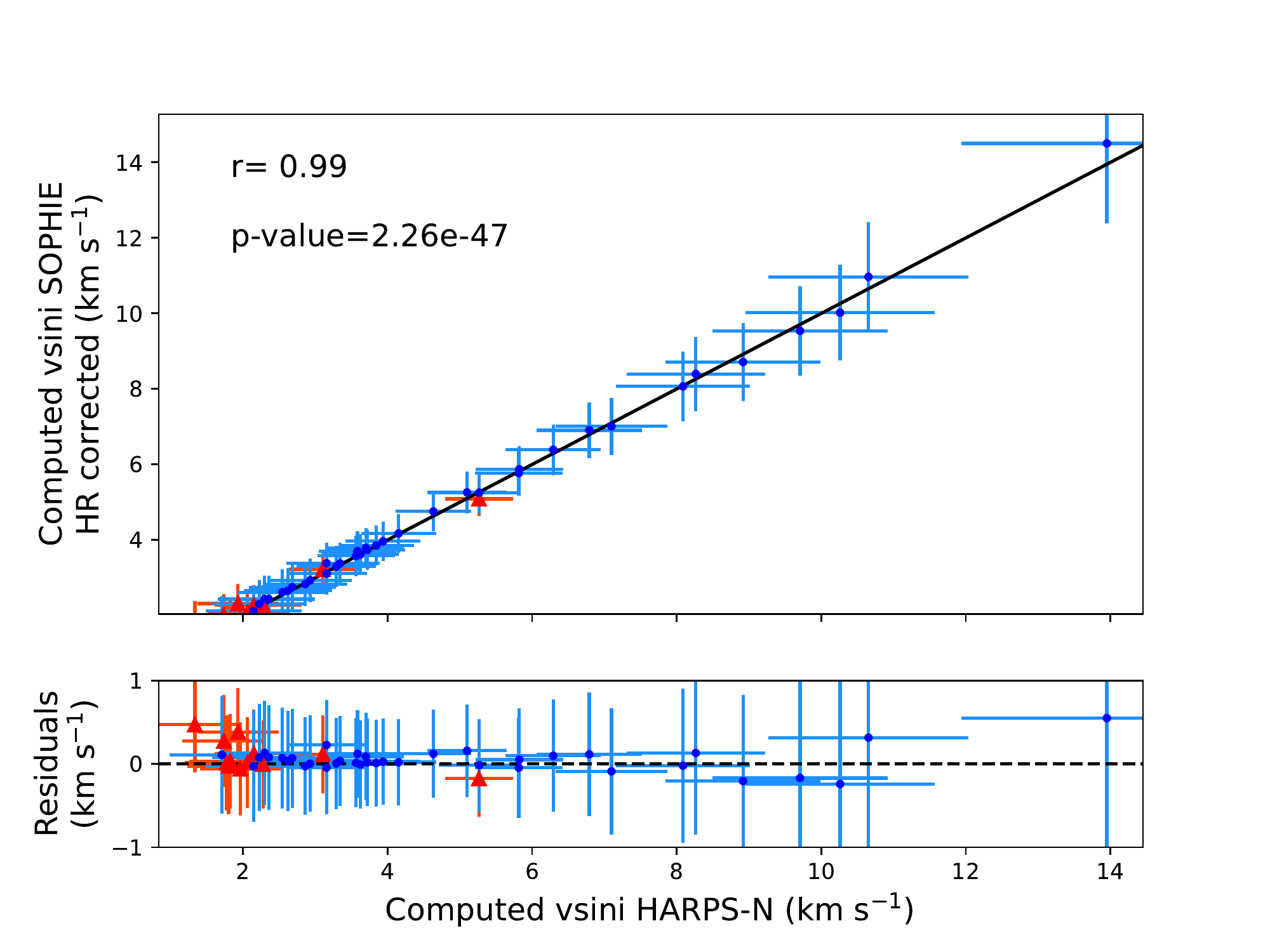}
    \caption{\textit{Upper panel:} comparison between $v_{\mathrm{eq}}\sin{i_\star}$ computed from the HARPS-N FWHM$_\mathrm{DRS}$ ($x$-axis) and those computed from the corrected SOPHIE high-resolution FWHM$_\mathrm{new}$ ($y$-axis). The blue dots are the values computed with the G2 mask relation, the red triangles are those computed with the K5 mask relation. The black line shows the one-to-one correlation. \textit{Lower panel:} residuals.}
    \label{fig:comparisonSOPHIEnew}
\end{figure}

The situation regarding the SOPHIE data is slightly different: applying the HARPS-N relation to the SOPHIE data results in \vsini\,\ values consistently overestimated, in particular at the lower end of the range.
This is not surprising, as the lower resolution of SOPHIE as compared to HARPS-N will result in larger FWHM$_\mathrm{DRS}$ values due to the greater instrumental broadening. Still, the effect is not simply a rigid shift, but it appears as a parabolic trend. We manipulated the SOPHIE FWHM$_\mathrm{DRS}$ values in order to correct them for the different instrumental resolution, using the following equation:
\begin{equation}
    \label{eq:correct_SOPHIE}
    \mathrm{FWHM}_\mathrm{new} = \sqrt{{\mathrm{FWHM}}_\mathrm{DRS}^2 - \left(\frac{c}{R_\mathrm{SOPHIE}}\right)^2 + \left(\frac{c}{R_\mathrm{HARPS-N}}\right)^2}
\end{equation}
where $c$ is the speed of light in \kms, and $R_\mathrm{SOPHIE}$ and $R_\mathrm{HARPS-N}$ are the resolution of SOPHIE and HARPS-N (see Table~\ref{tab:spectrographs}). The \vsini\,\ values computed with $\mathrm{FWHM}_\mathrm{new}$ are in much better agreement with those derived from HARPS-N data, as shown in Fig.~\ref{fig:comparisonSOPHIEnew}.
While the spread between HARPS-N and SOPHIE \vsini\,\ values is a bit larger than that between HARPS-N and HARPS ones, still it seems that our relation could be used also with the SOPHIE data, once they are corrected for the difference in resolution.

\begin{figure}
    \centering
    \includegraphics[width=\columnwidth]{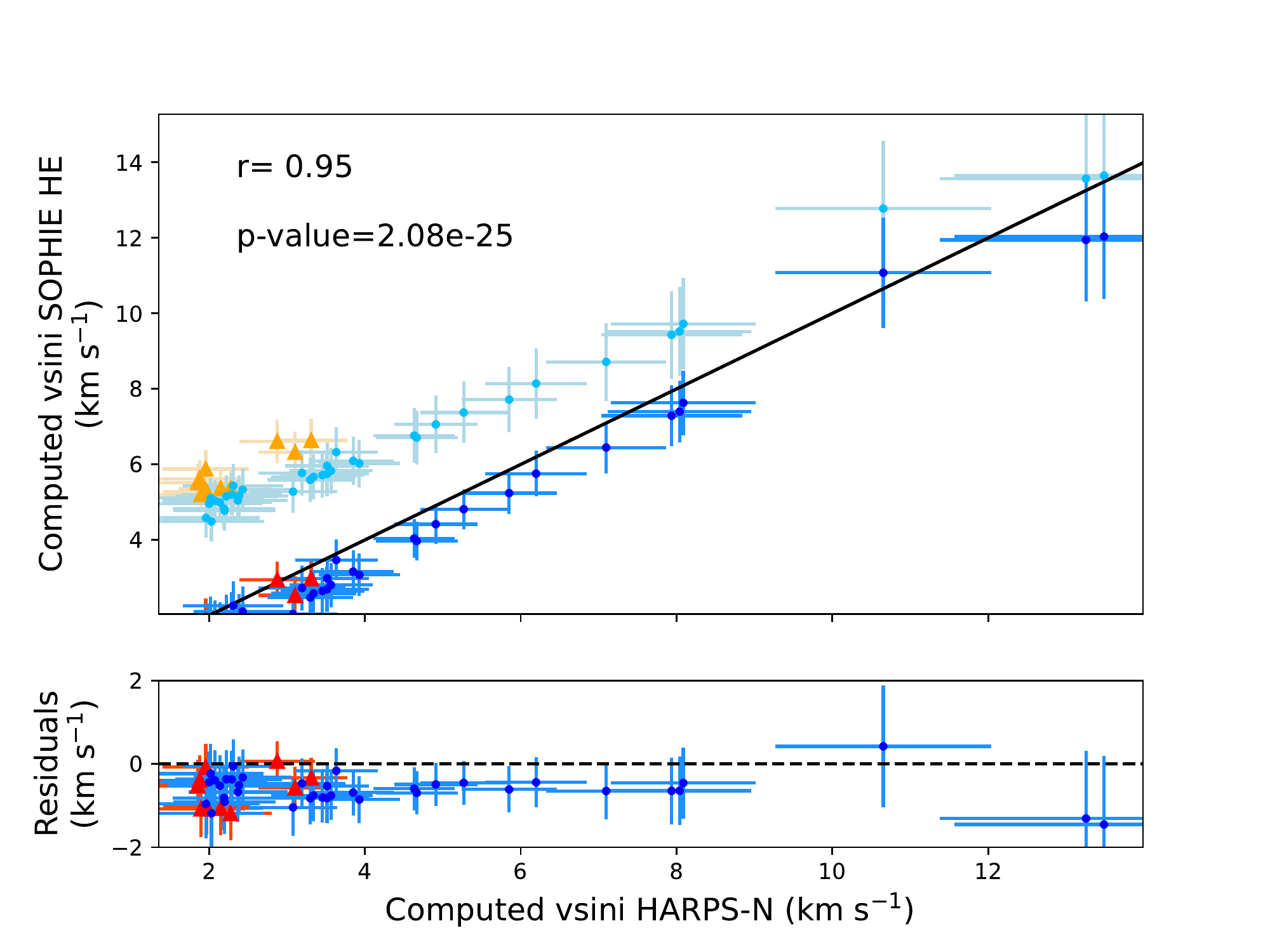}
    \caption{\textit{Upper panel:} comparison between $v_{\mathrm{eq}}\sin{i_\star}$ computed from the HARPS-N FWHM$_\mathrm{DRS}$ ($x$-axis) and those computed with SOPHIE in high-efficiency mode ($y$-axis). Orange triangles and light-blue dots are the results from SOPHIE high-efficiency FWHM$_\mathrm{DRS}$ with the K5 and G2 relation respectively, while the red triangles and blue dots are the results from the corrected SOPHIE high-efficiency FWHM$_\mathrm{new}$ ($y$-axis). The black line shows the one-to-one correlation. \textit{Lower panel:} residuals for the corrected SOPHIE high-efficiency FWHM$_\mathrm{new}$ only.}
    \label{fig:comparisonSOPHIEnewHE}
\end{figure}

To better test this assumption, we also selected the SOPHIE CCFs computed from the spectra observed in the HE mode and then we compared the \vsini\,\ computed from both the FWHM$_\mathrm{DRS}$ and FWHM$_\mathrm{new}$. The latter were derived using Eq.~\ref{eq:correct_SOPHIE} with the HE resolution. The results are shown in Fig.~\ref{fig:comparisonSOPHIEnewHE}: while correcting for the resolution does improve the agreement between HARPS-N and SOPHIE HE \vsini\,\ values, the results are still discrepant.
It seems then that a simple correction for the different resolution is not enough to adapt our relation to a different spectrograph, at least when the resolution difference is large enough. This assumes that there aren't any other factors in play, as, for example, a difference in the code to compute HR and HE CCFs in the SOPHIE DRS.

Unfortunately, we cannot test our method further on any other instrument, because very few spectrographs are equipped with dedicated DRS that yield also the CCFs in addition to the reduced spectra. ESPRESSO has the same capabilities (and a DRS derived from the HARPS one), but there are not enough public data from this instrument for a meaningful comparison. We were unable to compare the HARPS-N results with those obtained with instruments with a very different spectral coverage (such as the visible and near-infrared spectrograph CARMENES or the near-infrared spectrograph GIANO-B), because their DRSs do not compute any CCF.

Still, in case any other future DRS will yield also the CCFs, it will be of fundamental importance to calibrate or check and adapt this relation for each combination of instrument, wavelength range, spectral resolution, mathematical recipe (to compute both CCF and FWHM), and stellar mask. While the work is quite straightforward in the case of instruments such as HARPS-N (that offers a single, fixed choice of wavelength coverage and resolution), it may become slightly more complex when applied to instruments such as ESPRESSO (with 3 different resolving powers) or UVES (where a wide range of choices in both wavelength coverage and spectral resolution is available).

In any case, the strategy detailed in this paper in order to calibrate a FWHM$_\mathrm{DRS}$--$v_{\mathrm{eq}}\sin{i_\star}$ relation may be applied to any other relevant cases including self-made codes, allowing to better exploit the information carried in the CCFs.


\section{Conclusions} \label{sec:conclusions}
Using a well-defined set of calibrators, we were able to obtain two straightforward relations to obtain an estimation of the stellar $v_{\mathrm{eq}}\sin{i_\star}$ directly from the FWHM$_\mathrm{DRS}$ computed by the HARPS-N DRS using the G2 and K5 masks (see Eq.~\ref{eq:vsini}). These calibrations may be applied when the FWHM$_\mathrm{DRS}$ value is less than 20 km~s$^{-1}$. For larger values, other methods to compute the $v_{\mathrm{eq}}\sin{i_\star}$ are more accurate (i.e., Fourier transform or rotational profile fitting). Other relations were computed to be used when it is possible to estimate $v_{\rm micro}$ and/or $v_{\rm macro}$, and thus remove their contribution to the FWHM$_\mathrm{DRS}$.

We applied our basic relations to all the exoplanet-host stars found in the HARPS-N public archive and in the GAPS private data with CCF computed with G2 or K5 mask and FWHM$_\mathrm{DRS}$ < 20 km~s$^{-1}$: we obtained a catalog of homogeneous $v_{\mathrm{eq}}\sin{i_\star}$ measurements for 273 exoplanet-host stars. Of these stars, 206 have literature values of $v_{\mathrm{eq}}\sin{i_\star}$: comparing our results with those we found a very good agreement, with no object differing more than 3$\sigma$. Considering the stellar parameters when comparing our results with the literature, we constrain our relation to stars with $\log g > 3.5$.

We can reliably affirm that our simple FWHM$_\mathrm{DRS}$-$v_{\mathrm{eq}}\sin{i_\star}$ relations give solid results, comparable with those obtained with more sophisticated methods such as, for example, spectral synthesis. While our errors may overall be larger than those obtained in the literature, our results would still be useful in characterizing exoplanetary properties, and they may be used as a starting point for a more detailed analysis of the exoplanetary systems. In fact, we were able to determine or constrain the stellar inclination for 12 exoplanet-host stars with no previous $v_{\mathrm{eq}}\sin{i_\star}$ measurements, finding hints of spin-orbit misalignment in the K2-173 system.

We also tested our relations on the FWHM$_\mathrm{DRS}$ computed by the HARPS and SOPHIE DRS, and we conclude that Eq.~\ref{eq:vsini} may be used as it is also with HARPS data taken in the high accuracy mode ($R=115,000$). It would be possible to use our relation on the SOPHIE HR data once they are corrected for the different resolution, while using the SOPHIE HE data would require some additional fine-tuning. Still, the strategy detailed in this paper (selection of the calibration, creation of the FWHM$_\mathrm{DRS}$-$v_{\mathrm{eq}}\sin{i_\star}$ relation, test of the applicability range) may be used to calibrate other FWHM$_\mathrm{DRS}$-$v_{\mathrm{eq}}\sin{i_\star}$, with different combination of instrument resolutions, wavelength ranges, mathematical codes (to compute both CCF and FWHM), and stellar masks.

\begin{acknowledgements}
This paper is based on observations collected with the 3.58m Telescopio Nazionale Galileo (TNG), operated on
the island of La Palma (Spain) by the Fundaci\'{o}n Galileo Galilei of the INAF (Istituto Nazionale di Astrofisica) at the Spanish Observatorio del Roque de los Muchachos, in the frame of the programme Global Architecture of Planetary Systems (GAPS). This research used the facilities of the Italian Center for Astronomical Archive (IA2) operated by INAF at the Astronomical Observatory of Trieste. This research has made use of the SIMBAD database, operated at CDS, Strasbourg, France L.~M. acknowledges support from the ``Fondi di Ricerca Scientifica d'Ateneo 2021'' of the University of Rome ``Tor Vergata''. G.S. acknowledges support from CHEOPS ASI-INAF agreement n. 2019-29-HH.0.
\end{acknowledgements}

\bibliographystyle{aa} 
\bibliography{fwhm_vsini.bib} 

\end{document}